\documentclass[sigconf, nonacm]{acmart}

\usepackage{mathtools}
\usepackage{bm}
\usepackage{algorithm}
\usepackage{algpseudocodex}
\usepackage{subcaption}
\usepackage{cleveref}
\usepackage[acronym]{glossaries}
\usepackage{xcolor}
\usepackage{xspace}
\usepackage{multirow}

\theoremstyle{acmdefinition}
\newtheorem{remark}{Remark}

\DeclareMathOperator{\relation}{\mathcal{R}}
\DeclareMathOperator{\simrelation}{\preceq}

\newcommand{\reals}{{\ensuremath \mathbb{R}}}
\newcommand{\naturals}{{\ensuremath \mathbb{N}}}

\newcommand{\graph}{{\ensuremath \mathcal{G}}}
\newcommand{\graphverts}{{\ensuremath \mathcal{V}}}
\newcommand{\graphedges}{{\ensuremath \mathcal{E}}}
\DeclareMathOperator{\graphparent}{Pa}

\newcommand{\lipschitzconst}{{\ensuremath \mathcal{L}}}

\DeclareMathOperator*{\argmax}{argmax}

\newcommand{\uncertaintyspace}{{\ensuremath \Omega}}

\newcommand{\sigmaalgebra}{{\ensuremath \mathcal{F}}}
\newcommand{\borel}{{\ensuremath \mathcal{B}}}
\newcommand{\probmeas}{{\ensuremath \mathbb{P}}}
\newcommand{\auxprobmeas}{{\ensuremath \mathbb{Q}}}
\newcommand{\coupling}{{\ensuremath \mathbb{W}}}

\newcommand{\dists}{{\ensuremath \mathcal{D}}}
\DeclareMathOperator{\intamb}{int\,amb}

\newcommand{\ambiguityset}{{\ensuremath \Gamma}}
\newcommand{\probdist}{{\ensuremath \gamma}}

\newcommand{\randomvar}[1]{{\ensuremath \mathbf{#1}}}
\newcommand{\dataset}{{\ensuremath D}}
\newcommand{\sample}{{\ensuremath \tau}}
\newcommand{\confidenceparam}{{\ensuremath \beta}}
\newcommand{\numsamples}{{\ensuremath N}}

\newcommand{\region}{{\ensuremath q}}
\newcommand{\regionset}{{\ensuremath Q}}
\newcommand{\numregions}{{\ensuremath L}}
\newcommand{\auxregion}{{\ensuremath r}}

\newcommand{\numauxregions}{{\ensuremath J}}

\newcommand{\projection}{{\ensuremath \rho}}

\newcommand{\frs}{{\ensuremath F}}
\newcommand{\mytolerance}{{\ensuremath \varepsilon}}

\newcommand{\sysstate}{{\ensuremath x}}
\newcommand{\staterandomvar}{{\randomvar{\sysstate}}}
\newcommand{\initialstate}{{\ensuremath x_0}}
\newcommand{\noisebasic}{v}
\newcommand{\noise}{\randomvar{\noisebasic}}
\newcommand{\noisespace}{{\ensuremath V}}
\newcommand{\timevar}{{\ensuremath k}}

\newcommand{\control}{{\ensuremath u}}

\newcommand{\controller}{{\ensuremath \pi}}
\newcommand{\transitionkernel}{{\ensuremath T}}

\newcommand{\labelling}{{\ensuremath h}}
\newcommand{\outputspace}{{\ensuremath Y}}
\newcommand{\outputvar}{{\ensuremath y}}
\newcommand{\outputrandomvar}{\randomvar{\outputvar}}

\newcommand{\statespace}{{\ensuremath X}}
\newcommand{\abstractionmap}{{\ensuremath \mathcal{T}}}
\newcommand{\refinement}{{\ensuremath \mathcal{I}}}
\newcommand{\actionmap}{{\ensuremath \mathcal{A}}}

\newcommand{\initialset}{{\ensuremath \statespace_0}}
\newcommand{\controlspace}{{\ensuremath U}}

\newcommand{\source}{{\ensuremath s}}
\newcommand{\initialsource}{{\ensuremath s_0}}
\newcommand{\destination}{{\ensuremath t}}

\newcommand{\action}{{\ensuremath a}}
\newcommand{\stateset}{{\ensuremath S}}

\newcommand{\actionset}{{\ensuremath A}}
\newcommand{\scheduler}{{\ensuremath \sigma}}
\newcommand{\schedulerset}{{\ensuremath \Pi}}

\newcommand{\adversary}{{\ensuremath \nu}}

\newcommand{\initialstateset}{{\ensuremath \stateset_0}}
\newcommand{\specification}{{\ensuremath \varphi}}

\newcommand{\automaton}{{\ensuremath \mathcal{A}}}

\newcommand{\alphabet}{{\ensuremath \Sigma}}
\newcommand{\automatonstate}{{\ensuremath z}}
\newcommand{\automatonstatespace}{{\ensuremath Z}}
\newcommand{\automatontransition}{{\ensuremath \delta}}
\newcommand{\omegaword}{{\ensuremath \mathsf{w}}}
\newcommand{\omegaletter}{{\ensuremath w}}
\newcommand{\dpapriority}{{\ensuremath \mathsf{pr}}}

\newacronym{pdf}{PDF}{probability density function}
\newacronym{iid}{i.i.d.}{independent and identically distributed}
\newacronym{pwa}{PWA}{piecewise affine}
\newacronym{pwc}{PWC}{piecewise constant}
\newacronym{mbr}{MBR}{Minimum Bounding Rectangle}
\newacronym{wrt}{wrt.}{with respect to}
\newacronym{wlog}{wlog.}{without loss of generality}

\newacronym{scc}{SCC}{Strongly Connected Component}
\newacronym{bscc}{BSCC}{Bottom \acrshort{scc}}

\newacronym{lp}{LP}{Linear Programming}
\newacronym{gd}{GD}{Gradient Descent}
\newacronym{dp}{DP}{Dynamic Programming}
\newacronym{sdp}{SDP}{Semi-definite Programming}
\newacronym{sos}{SoS}{Sum-of-Squares}
\newacronym{sosp}{SoSP}{Sum-of-Squares Programming}
\newacronym{pac}{PAC}{Probably Approximately Correct}
\newacronym{saa}{SAA}{Sample Average Approximation}

\newacronym{nndm}{NNDM}{Neural Network Dynamic Model}
\newacronym[longplural=Gaussian Processes]{gp}{GP}{Gaussian Process}
\newacronym{nn}{NN}{Neural Network}
\newacronym{bnn}{BNN}{Bayesian Neural Network}

\newacronym{ltl}{LTL}{Linear Temporal Logic}
\newacronym{scltl}{scLTL}{syntatically co-safe Linear Temporal Logic}
\newacronym{lbp}{LBP}{Linear Bound Propagation}
\newacronym{ibp}{IBP}{Interval Bound Propagation}
\newacronym{sbf}{SBF}{Stochastic Barrier Function}
\newacronym{nbf}{NBF}{Neural Barrier Function}
\newacronym{pw-sbf}{PW-SBF}{piecewise SBF}
\newacronym{pwc-sbf}{PWC-SBF}{piecewise constant SBF}
\newacronym{pwa-sbf}{PWA-SBF}{piecewise affine SBF}
\newacronym[longplural=Interval Markov Decision Processes]{imdp}{IMDP}{Interval Markov Decision Process}
\newacronym[longplural=Robust Markov Decision Processes]{rmdp}{RMDP}{Robust Markov Decision Process}
\newacronym[longplural=Markov Decision Processes]{mdp}{MDP}{Markov Decision Process}
\newacronym[longplural=factored Markov Decision Processes]{fmdp}{fMDP}{factored Markov Decision Process}
\newacronym{dbn}{DBN}{Dynamic Bayesian Network}
\newacronym{dcn}{DCN}{Dynamic Credal Network}
\newacronym{imc}{IMC}{Interval Markov Chain}
\newacronym[longplural=orthogonally decoupled Interval Markov Decision Processes]{odimdp}{odIMDP}{orthogonally decoupled Interval Markov Decision Process}
\newacronym[longplural=factored Interval Markov Decision Processes]{fimdp}{fIMDP}{factored Interval Markov Decision Process}
\newacronym{odimc}{odIMC}{orthogonally decoupled Interval Markov Chain}
\newacronym{cegis}{CEGIS}{Counter-Example Guided Inductive Synthesis}

\newacronym{mpc}{MPC}{Model Predictive Control}

\newacronym{gpu}{GPU}{General-Purpose Graphics Processing Unit}
\newacronym{csc}{CSC}{Compressed Sparse Column}
\newacronym{simd}{SIMD}{Single Instruction Multiple Data}
\newacronym{simt}{SIMT}{Single Instruction Multiple Thread}
\newacronym{api}{API}{Application Programming Interface}
\newacronym{add}{add}{Algebraic Decision Diagram}
\newacronym[longplural=Deterministic Finite Automata]{dfa}{DFA}{Deterministic Finite Automaton}
\newacronym[longplural=Deterministic Parity Automata]{dpa}{DPA}{Deterministic Parity Automaton}

\newacronym{adas}{ADAS}{Advanced Driver Assistance System}
\newacronym{alks}{ALKS}{Automated Lane Keeping System}
\newacronym{aebs}{AEBS}{Automated Emergency Braking System}
\newacronym{vmt}{VMT}{Vehicle Miles Travelled}

\newacronym{cps}{CPS}{Cyber-Physical System}

\renewcommand\footnotetextcopyrightpermission[1]{} %
\newcommand{\mc}[1]{\mathcal{#1}}

\newtheorem{problem}{Problem}
\newtheorem{assumption}{Assumption}

\title[Data-Driven Factored Abstractions of Stochastic Systems]{Data-Driven Factored Abstractions of Stochastic Systems}

\begin{document}

\author{Mahdi Nazeri}
\authornote{Equal contribution}
\affiliation{%
  \institution{University of Oxford \& MPI SWS}
  \city{Oxford}
  \country{United Kingdom}
}
\email{mahdi.nazeri@cs.ox.ac.uk}

\author{Frederik Baymler Mathiesen}
\authornotemark[1]
\affiliation{%
  \institution{University of Oxford}
  \city{Oxford}
  \country{United Kingdom}
}
\email{frederik.mathiesen@cs.ox.ac.uk}

\author{Sadegh Soudjani}
\affiliation{%
  \institution{University of Birmingham \& MPI SWS}
  \city{Birmingham}
  \country{United Kingdom}
}
\email{s.soudjani@bham.ac.uk}

\author{Alessandro Abate}
\affiliation{%
  \institution{University of Oxford}
  \city{Oxford}
  \country{United Kingdom}
}
\email{alessandro.abate@cs.ox.ac.uk}

\begin{abstract}
Formal verification and correct-by-design policy synthesis for stochastic dynamical systems face significant scalability challenges due to the curse of dimensionality. These challenges are exacerbated when the system dynamics are unknown and abstractions must be constructed from data. This paper introduces a data-driven framework for constructing \emph{factored} interval Markov decision process (fIMDP) abstractions from noisy transition observations. Assuming conditional independence of successor coordinates and known coordinate-wise Lipschitz bounds on the dynamics, the framework estimates marginal transition probabilities rather than joint probabilities, resulting in an exponential reduction in the number of probability entries to be estimated and stored.
We establish finite-sample confidence guarantees that the abstraction captures the concrete transition probabilities uniformly over each abstract region, enabling certified lower bounds on satisfaction probabilities for $\omega$-regular specifications. 
Experiments on benchmark systems, including pendulum swing-up, unicycle control, and a four-dimensional planar robot model, demonstrate improvements in certified satisfaction probability lower bounds of up to 25 percentage points compared with a state-of-the-art flat IMDP abstraction method. Our framework required up to 23x less memory and was consistently faster in constructing the abstraction.
\end{abstract}



\maketitle

\section{Introduction}
\label{sec:intro}

Stochastic dynamical systems operating in uncertain environments require rigorous guarantees on safety and task satisfaction~\cite{alur2015cps,baier2008principles}.
Formal abstractions support verification and correct-by-design policy synthesis by representing continuous-state dynamics through finite-state models with quantified guarantees.
When the dynamics or noise distributions are unknown, constructing such abstractions requires reasoning from observations while accounting for the uncertainty introduced by finite data.
Data-driven formal abstraction methods provide a principled basis for this task; see, e.g.,~\cite{DBLP:conf/hybrid/JacksonLFL21,kazemi2024data,nazeri2025data} and the recent survey~\cite{samari2026data}.

Despite recent progress, scaling data-driven abstractions to high-dimensional state spaces remains challenging.
In a flat abstraction, (bounds on) the transition probability must be estimated and stored for each abstract state and action over the full abstract state and action space, leading to many probabilities to estimate, which in turn requires much data, and substantial storage requirements.
Factored representations exploit conditional independence between successor-state components to represent these transitions through marginal distributions.
For continuous-state stochastic systems, this structure has been used to construct formal abstractions represented as dynamic Bayesian networks from a given model~\cite{SAM2015DBN,SAM2017DBN}.
Subsequent developments have exploited structured representations for abstraction-based control synthesis~\cite{van2023syscore,mathiesen2025scalable,wang2026correct}.
These results motivate the construction of factored abstractions from observations when the underlying dynamics and noise distributions are unknown.

Learning robust factored models from transition observations has been studied for finite \glspl{mdp}~\cite{schnitzer2026efficient}.
In this paper, we construct factored abstractions of continuous-state systems, combining Lipschitz-based reachable-set enclosures with marginal transition counts to obtain probability bounds valid throughout each abstract region with a prescribed confidence.
A theoretical analysis of the sample complexity shows that the factored representation is significantly more sample efficient than an equivalent flat model, due to requiring to estimate fewer transition probability bounds.

We evaluate the framework on five stochastic dynamical systems: a double integrator, a pendulum swing-up problem, a mountain car, a unicycle, and a four-dimensional planar robot with friction.
We assess the certified satisfaction probability lower bounds, abstraction construction time, and storage requirements compared with a flat abstraction to \glspl{imdp}~\cite{nazeri2025data2}.
The experiments demonstrate how exploiting factorization yields more compact abstractions and tighter bounds on the satisfaction probability using the same transition data and overall confidence level.

The main contributions of this work are:
\begin{itemize}
    \item A data-driven algorithm for constructing \glspl{fimdp} from noisy transition observations, using reachable-set enclosures and marginal transition counts.
    \item Finite-sample confidence guarantees that the abstraction captures the concrete transition probabilities uniformly over
    each abstract region, enabling the lifting of satisfaction probability lower bounds over the abstract model to the concrete system.
    \item An analysis of the reduction in stored transition probability bounds and the resulting allocation of the statistical confidence budget, together with an experimental comparison against flat \gls{imdp} abstractions.
\end{itemize}

\paragraph{Paper Organization.} The remainder of the paper is organized as follows.
\Cref{sec:related} discusses related work.
\Cref{sec:prelim} introduces \glspl{imdp} and their factored representations, and \Cref{sec:problem} states the assumptions and synthesis problem.
\Cref{sec:framework} presents the data-driven abstraction construction and its guarantees.
\Cref{sec:experiments} reports the experimental evaluation, and \Cref{sec:conclusion} concludes the paper.

\section{Related Work}
\label{sec:related}

Finite-state abstractions of continuous-state stochastic systems build on discretization schemes for stochastic dynamic
programming~\cite{bertsekas1975convergence}.
Their use in formal verification was developed together with quantitative bounds on the resulting approximation error~\cite{abate2008probabilistic,abate2010approximate}.
Subsequent approaches have improved these bounds through adaptive gridding~\cite{esmaeil2013adaptive} and coupling-based constructions for linear stochastic systems~\cite{vanhuijegevoort2022similarity}.
An alternative to bounding the approximation error separately is to incorporate variation within each abstract region directly into the transition model.
This leads to \gls{imdp} abstractions, where intervals bound the probabilities of reaching each destination region from the concrete states represented by an abstract state~\cite{lahijanian2015formal,cauchi2019efficiency}.
Such constructions have been developed for linear systems~\cite{cauchi2019efficiency} and nonlinear or switched stochastic systems~\cite{lahijanian2015formal,dutreix2020interval,adams2022formal}.

Factored representations have a long history in probabilistic planning and structured \glspl{mdp}.
By describing states through variables and transitions through conditional probability distributions, dynamic Bayesian networks can represent large transition models compactly~\cite{boutilier1999decision,koller1999computing}.
Compact transition representations do not, however, automatically yield compact value functions or tractable planning algorithms~\cite{koller1999computing}.
Factored representations have also been extended to \glspl{mdp} with imprecise transition probabilities, together with methods for robust policy computation~\cite{delgado2009representing}.

For continuous-state stochastic systems, dynamic Bayesian networks have been used to represent finite-state abstractions by exploiting independence between noise components and dependencies between state variables~\cite{SAM2015DBN,SAM2017DBN}.
These constructions compute coordinate-wise transition distributions from a known model and establish quantitative error bounds for finite-horizon probabilistic invariance.
More recent approaches exploit structured representations for stochastic control synthesis.
In particular, additive noise structure has also been exploited to separate the noisy and controlled Bellman updates~\cite{van2022temporal,van2023syscore}, while structural \gls{imdp} abstractions and tensor-based representations are investigated in~\cite{mathiesen2025scalable,wang2026correct}.
Probabilistic simulation relations provide a means of transferring guarantees between abstract and concrete systems \cite{doi:10.1137/16M1079397}.

Compositional approaches address scalability by constructing abstractions for individual subsystems and combining their
guarantees~\cite{Lavaei2020c}.
The combination can be justified through small-gain conditions~\cite{Lavaei2018b,Lavaei2019c},
dissipativity arguments~\cite{Lavaei2018,Lavaei2020}, or sub-simulation relations~\cite{schon2023verifying}.
These methods exploit a subsystem decomposition, whereas factored representations retain a joint abstract state space while representing its transition structure compactly.
The applicability and computational benefits of compositional methods depend on the subsystem dimensions, interconnection conditions, and the specifications supported by the chosen composition rule.

Data-driven abstraction methods construct finite-state models from observations, reducing the need for an explicit description of the dynamics or noise distribution.
The central challenge is to account for statistical uncertainty while establishing guarantees that hold throughout each abstract region.
Gaussian process-based approaches learn a model of the unknown dynamics and use uncertainty bounds to support verification and abstraction construction~\cite{jackson2020safety,jiang2022safe,reed2025error,schon2024bayesian,DBLP:conf/hybrid/JacksonLFL21,schon2024data}.
These guarantees depend on assumptions about the unknown function, noise, and regression model.
Obtaining valid uncertainty bounds when the kernel hyperparameters are unknown is itself a significant challenge~\cite{capone2022gaussian}.

Another line of work constructs data-driven abstractions with probabilistic closeness guarantees between abstract and concrete trajectories~\cite{zhang2024formal,DBLP:journals/csysl/LavaeiSFZ23}.
Such guarantees support the transfer of verification and synthesis results between the two models.
However, bounds that accumulate approximation error over time can become conservative for long-horizon specifications.
This motivates abstraction constructions that incorporate uncertainty directly into the transition model and support robust verification and synthesis.

Robust \gls{mdp} abstractions incorporate uncertainty in the sets of admissible transition distributions.
When these sets contain the concrete transition probabilities uniformly over each abstract region with high confidence, robust synthesis provides statistically certified guarantees
for the concrete system.
For known linear dynamics with unknown additive noise, \gls{imdp} abstractions can be constructed using scenario optimization over noise samples~\cite{badings2023robust}.
Wasserstein ambiguity sets have also been used to construct robust abstractions for nonlinear systems with more general disturbances~\cite{gracia2024data}.
Extensions address unknown nonlinear dynamics using noise-free observations of the dynamics~\cite{gracia2024temporal}.
Clopper-Pearson confidence intervals provide an alternative construction of \gls{imdp} abstractions for unknown nonlinear systems~\cite{nazeri2025data}.
More recently, \gls{imdp} abstractions have been constructed directly from noisy transition observations, accounting for variation within each abstract region through reachable-set enclosures~\cite{nazeri2025data2}.

Robust factored \glspl{mdp} can also be learned from transition observations, with marginal uncertainty sets supporting high-confidence policy performance guarantees~\cite{schnitzer2026efficient}.
The approach of \cite{schnitzer2026efficient} considers an unknown finite-state system.
Our construction instead abstracts a continuous-state system, requiring transition bounds that account for both statistical uncertainty and variation across every concrete state within each abstract region.

These developments provide complementary foundations for data-driven abstraction and robust planning.
Our contribution is a construction of factored \gls{imdp} abstractions for unknown continuous-state stochastic systems from noisy transition observations.
Under conditional independence of successor coordinates and known coordinate-wise Lipschitz bounds, the construction provides simultaneous confidence guarantees on marginal transition bounds that hold throughout each abstract region.
Factorization reduces the number of probability bounds to estimate and store, while preserving the guarantees needed for robust policy synthesis.

\section{Preliminaries}
\label{sec:prelim}
We denote by $\reals$ and $\naturals$ the sets of real numbers and non-negative integers, respectively. Let $\projection_l : \reals^n \to \reals$, $y \mapsto y_l$, denote the projection onto the $l$-th coordinate. A probability space $(\uncertaintyspace, \sigmaalgebra, \probmeas)$ consists of a sample space $\uncertaintyspace$, a $\sigma$-algebra $\sigmaalgebra$, and a probability measure $\probmeas : \sigmaalgebra \to [0, 1]$. $\borel(\statespace)$ denotes the Borel $\sigma$-algebra on the set $\statespace$. A random vector $\noise$ is a measureable function $\noise : \uncertaintyspace \to \reals^n$, $n \in \naturals$. We denote by $\dists(\statespace)$ the set of probability distributions over a set $\statespace$. An interval ambiguity set $\ambiguityset \subseteq \dists(\statespace)$ is a subset of distributions defined by two bounding functions $\underline{\probdist}$, $\overline{\probdist}$ that satisfy $\underline{\probdist}(x) \leq \overline{\probdist}(x)$ for all $x \in \statespace$ and $\sum_{x \in \statespace} \underline{\probdist}(x) \leq 1 \leq \sum_{x \in \statespace} \overline{\probdist}(x)$. In particular, $\ambiguityset = \{\gamma \in \dists(\statespace) \,:\, \underline{\probdist}(x) \leq \probdist(x) \leq \overline{\probdist}(x), \forall x \in \statespace \}$. We denote by $\intamb(\statespace)$ the set of all interval ambiguity sets on $\statespace$. 

\subsection{\texorpdfstring{$\omega$}{omega}-regular Specifications}
Let $\outputspace$ be a finite set of atomic propositions and $\alphabet = 2^{\outputspace}$ the alphabet.
An infinite word is a sequence $\omegaword = \omegaletter_0 \omegaletter_1 \ldots \in \alphabet^\omega$.
An $\omega$-regular property is a language $\specification \subseteq \alphabet^\omega$, which is recognised by a \gls{dpa}~\cite{10.1109/SFCS.1991.185392}.

\begin{definition}[DPA]
A \acrfull{dpa} is a tuple $\automaton = (\automatonstatespace, \alphabet, \automatontransition, \automatonstate_0, \dpapriority)$, where $\automatonstatespace$ is a finite set of states, $\automatontransition : \automatonstatespace \times \alphabet \to \automatonstatespace$ is a total transition function, $\automatonstate_0 \in \automatonstatespace$ is the initial state, and $\dpapriority : \automatonstatespace \to \naturals$ is a priority function.
\end{definition}

A word $\omegaword$ induces a unique run $\rho = \automatonstate_0 \automatonstate_1 \ldots$ with $\automatonstate_{i+1} = \automatontransition(\automatonstate_i, \omegaletter_i)$.
Let $\mathrm{Inf}(\rho)$ denote the set of states visited infinitely often by $\rho$.
The run is accepting if $\min\{\dpapriority(\automatonstate) : \automatonstate \in \mathrm{Inf}(\rho)\}$ is even,
and the language $\mathcal{L}(\automaton)$ of $\automaton$ is the set of words with an accepting run.
Every $\omega$-regular property $\specification$ is recognised by some \gls{dpa}~\cite{10.1109/SFCS.1991.185392}.
Given a labelling function $\labelling : \stateset \to \alphabet$, a path $\source_0 \source_1 \ldots$ satisfies $\specification$ if $\labelling(\source_0) \labelling(\source_1) \ldots \in \mathcal{L}(\automaton_\specification)$.

\subsection{Interval Markov Decision Processes}
\Acrfullpl{imdp} are a generalization of \glspl{mdp} where, rather than assuming perfect knowledge of the transition probabilities, they are bounded by intervals \cite{givan2000bounded}. In other words, an \gls{imdp} represents an (uncountable) family of \glspl{mdp}. 
\begin{definition}[Interval Markov Decision Process]
An \acrfull{imdp} is a tuple $\mc{M} = (\stateset, \actionset, \ambiguityset)$, where:
\begin{itemize}
    \item $\stateset$ is a finite set of states;
    \item $\actionset$ is a finite set of actions; and
    \item $\ambiguityset = \{\ambiguityset_{\source, \action}\}_{\source \in \stateset, \action \in \actionset}$ is a collection of interval ambiguity sets, $\ambiguityset_{\source, \action} \in \intamb(\stateset)$, indexed by the states $\source \in \stateset$ and actions $\action \in \actionset$.
\end{itemize}
\end{definition}
Although more general models are available, \glspl{imdp} are attractive because robust value iteration over this class can be efficiently computed via the O-maximization algorithm \cite{givan2000bounded, lahijanian2015formal}.

\subsection{Factored Representation}\label{subsec:fimdp}
An issue with the \emph{flat} structure of \glspl{imdp} and \glspl{mdp} is the space complexity of the transition function/ambiguity sets; it creates a practical bottleneck in the fidelity of abstractions. Instead, recent research has shown that factored models where the transition function is decomposed into state variables can greatly reduce this complexity \cite{van2023syscore, mathiesen2025scalable, wang2026correct}. 
\begin{definition}[{\Acrfull{fimdp}}]\label{def:fimdp}
    Let $\stateset_1, \ldots, \stateset_n$ be finite sets of values that each state variable $\stateset_i \in \{\source^i_1, \ldots, \source^i_{n_i}\}$ may take on. Similarly, define the finite sets of values $\actionset_1, \ldots, \actionset_m$ for each action variable $\actionset_j \in \{\action^j_1, \ldots, \action^j_{n_j}\}$. Furthermore, assume a given bipartite dependency graph $\graph = (\graphverts, \graphedges)$ with \begin{equation}
        \graphverts = \underbrace{\{\source^1, \ldots, \source^n\} \cup \{\action^1, \ldots, \action^m\}}_{\text{Condition variables}} \;\, \cup \underbrace{\{\stateset_1, \ldots, \stateset_n\}}_{\text{Target variables}}.
    \end{equation}
    Let $\graphparent_{\graph}(\stateset_i)$ denote the parent variables of $\stateset_i$ in the graph $\graph$. In other words, $\graph$ is a dependency graph and $\graphparent_\graph(\stateset_i)$ denotes the conditional variables for the state variable $\stateset_i$.
    Then, an \gls{fimdp} is a tuple $M = (\stateset, \actionset, \graph, \ambiguityset)$ where
    \begin{itemize}
        \item $\stateset = \stateset_1 \times \cdots \times \stateset_n$ is a finite set of joint states;
        \item $\actionset = \actionset_1, \ldots, \actionset_m$ is a finite set of joint actions; and
        \item $\ambiguityset = \{\ambiguityset_{\source, \action}\}_{\source \in \stateset, \action \in \actionset}$ are sets of feasible transition probability distributions where $\ambiguityset_{\source, \action} = \bigotimes_{i=1}^n \ambiguityset^i_{(\source, \action) \cap \graphparent_\graph(\stateset_i)}$ with each marginal $\ambiguityset^i_{(\source, \action) \cap \graphparent_\graph(\stateset_i)} \in \intamb(\stateset_i)$.
    \end{itemize}
\end{definition}
For the purpose of this manuscript, we assume that the dependency graph will be a complete bipartite graph.
Thus, for brevity, we write unambiguously $\ambiguityset^i_{\source, \action}$ to mean $\ambiguityset^i_{(\source, \action) \cap \graphparent_\graph(\stateset_i)}$. We use $\scheduler$ to denote a scheduler for the \gls{fimdp}. Note that, since we are considering $\omega$-regular objectives (see \Cref{prob:main_data-driven}), it is necessary that the scheduler carries (finite) memory \cite{baier2008principles}. However, since memory construction is standard (product between specification automaton and system model), we refer to \cite{baier2008principles,cauchi2019efficiency,9144391} for details. We denote the set of schedulers for a model $M$ by $\schedulerset_M$.

A scheduler alone does not determine a probability measure on an \gls{fimdp}, since its transition distributions are uncertain.
An environment strategy $\adversary$ resolves this uncertainty by selecting a distribution from $\ambiguityset_{\source,\action}$ after each action, possibly depending on the finite history.
For a scheduler $\scheduler$ and initial state $\source$, we define the robust, or pessimistic, satisfaction probability as
\begin{equation}
    \probmeas^{M,\scheduler}_{\source}(\specification)
    \coloneqq
    \inf_{\nu}
    \probmeas^{M,\scheduler,\adversary}_{\source}(\specification),
\end{equation}
where the probability on the right is induced jointly by $\scheduler$ and $\adversary$, and $\specification$ is an $\omega$-regular specification on labelled paths.
The environment may choose a new admissible distribution at every step.

Despite this seemingly more complex model structure, the storage requirements are exponentially reduced relative to flat \glspl{imdp}. Consequently, we would like to extend data-driven abstraction techniques to this model class.

\section{Problem Statement}
\label{sec:problem}

A stochastic dynamical system $\mc{M}$ is defined over discrete steps $\timevar \in \naturals$ by a stochastic difference equation
\begin{equation}\label{eq:system}
    \staterandomvar[\timevar+1] = f(\staterandomvar[\timevar], \control[\timevar], \noise[\timevar]), \quad \outputrandomvar[\timevar] = \labelling(\staterandomvar[\timevar])
\end{equation}
where $\staterandomvar[\timevar]$ is the state of the system taking values in $\statespace \subseteq \reals^n$, with starting from a state in a set of initial states $\staterandomvar[0] = \initialstate \in \initialset$, and $\control[\timevar] \in \controlspace \subseteq \reals^m$ is the control input.
The process noise $\{\noise[\timevar]\}_{\timevar \in \naturals}$ is a sequence of \gls{iid} random variables $\noise[\timevar] : \uncertaintyspace \to \noisespace$ defined on the probability space $(\uncertaintyspace, \sigmaalgebra, \probmeas)$. 

The control inputs are chosen by a policy $\controller = (\controller[0], \controller[1], \ldots)$, where each $\controller[\timevar]$ is a measurable map from the finite state--control history $(\sysstate[0], \control[0], \ldots,  \sysstate[\timevar - 1], \control[\timevar - 1], \sysstate[\timevar])$ to $\controlspace$ \cite{bertsekas1978stochastic,puterman1994markov}.
This policy class includes finite-memory controllers obtained by combining state feedback with a specification automaton. A given initial state and policy induce a probability measure $\probmeas^{\controller,\initialstate}_{\mc{M}}$ on the stochastic process $\{\staterandomvar[\timevar]\}_{\timevar\in\naturals}$ \cite{baier2008principles,schon2024bayesian}.

To enable a factored abstraction, we assume that each output of $f$ has independent noise \cite{SAM2015DBN,SAM2017DBN}.
\begin{assumption}[Independent noise]\label{ass:independent_noise}
    For every $i \in \{1,\ldots,n\}$, the dynamics satisfy $\staterandomvar_i[\timevar+1] = f_i\left(\staterandomvar[\timevar], \control[\timevar], \noise_i[\timevar]\right)$.
    At each time step $\timevar$, the noise components $\noise_1[\timevar],\ldots,\noise_n[\timevar]$ are mutually independent.
    The noise vectors $\noise[\timevar]$ are independent and identically distributed across time, and each $\noise[\timevar]$ is independent of the initial state and the state and control history up to time $\timevar$.
\end{assumption}

\begin{remark}
    Assumption~\ref{ass:independent_noise} concerns independence between noise components and does not require their dependence on the state and control input to be additive or multiplicative.
    This assumption naturally holds when different state coordinates are affected by independent sources of uncertainty.
    In some cases, a suitable change of state coordinates can also yield the required structure.
\end{remark}

Associated with the dynamics in~\eqref{eq:system} is a transition kernel $\transitionkernel(\cdot \mid \sysstate,\control)$, where $\transitionkernel(B \mid \sysstate,\control)$ denotes the probability that the successor state belongs to a Borel set $B \subseteq \reals^n$, given the current state $\sysstate$ and control input $\control$.
Under Assumption~\ref{ass:independent_noise}, this kernel factorizes as
\begin{equation}\label{eq:decomposed_transition_kernel}
    \transitionkernel(B \mid \sysstate,\control) = \prod_{i=1}^n \transitionkernel_i(B_i \mid \sysstate,\control), \qquad B = \prod_{i=1}^n B_i,
\end{equation}
where each $B_i \subseteq \reals$ is Borel measurable and $\transitionkernel_i$ is the $i$-th marginal transition kernel.
In Appendix~\ref{app:decomposition}, we prove that for a given grid size, the independence between marginals in \eqref{eq:decomposed_transition_kernel} is necessary \emph{on the grid} to build the product form ambiguity sets.

We assume that both $f$ and $\probmeas$ are unknown.
In place of knowing $f$ and $\probmeas$, we rely on data and Lipschitz continuity. In particular, we assume that each $f_i : \statespace \times \controlspace \times \uncertaintyspace \to \reals$ is Lipschitz continuous in the state and control input with known constants $\lipschitzconst^i_\statespace, \lipschitzconst^i_\controlspace$. 
Note that $\lipschitzconst^i_\statespace$ and $\lipschitzconst^i_\controlspace$ can be upper bounds; not necessarily the smallest constants to satisfy the inequality.
\begin{definition}[Element-wise Lipschitz]
    System $\mc{M}$ is \emph{element-wise} Lipschitz continuous in $\sysstate$ and $\control$ if for every dimension $i \in \{1, \ldots, n\}$ there exists constants $\lipschitzconst^i_\statespace, \lipschitzconst^i_\controlspace \in \reals_{\geq 0}$ such that for every $\noisebasic \in \noisespace$
    \begin{equation}\label{eq:lipschitz}
        \begin{aligned}
            &\lvert f_i(\sysstate_1, \control_1, \noisebasic) - f_i(\sysstate_2, \control_2, \noisebasic) \rvert \leq \lipschitzconst^i_\statespace \lVert \sysstate_1 - \sysstate_2 \rVert + \lipschitzconst^i_\controlspace \lVert \control_1 - \control_2 \rVert
        \end{aligned}
    \end{equation}
    for all $\sysstate_1, \sysstate_2 \in \statespace$ and $\control_1, \control_2 \in \controlspace$,
    where $\lVert \cdot \rVert$ is the Euclidean norm. 
\end{definition}

The data from the system are samples of the form $\sample_j = (\sysstate_j, \control_j, \sysstate^+_j)$ such that $\sysstate^+_j = f(\sysstate_j, \control_j, \noisebasic_j)$, that is, we do not have access to the underlying noise sample $\noisebasic_j$. 
\begin{definition}[Dataset]
    A dataset $\dataset$ is a collection of $\numsamples$ samples $\dataset = \{\sample_1, \ldots, \sample_\numsamples\}$. 
    We assume that all samples in $\dataset$ are \gls{iid} with an underlying probability measure $\auxprobmeas$.
\end{definition}
The underlying distribution implies that the dataset is sampled according to the product measure $\auxprobmeas^\numsamples$. 

The goal is to compute the probability that the system, with limited information, satisfies an $\omega$-regular specification $\specification$ where the objectives are given by a labelling function $\labelling : \statespace \to 2^\outputspace$, which maps each state to a finite set of labels from the (finite) output space $\outputspace$.
Formally, we consider the following problem.

\begin{problem}[Verification]\label{prob:main_data-driven}
Given an unknown stochastic system $\mc{M}$, an initial state $\initialstate \in \statespace$, an $\omega$-regular specification $\specification$, a labelling function $\labelling$, a dataset $\dataset = \{\sample_1, \ldots, \sample_\numsamples\}$, and a confidence budget $\confidenceparam \in (0,1)$, construct an abstract model $\hat{\mc{M}}$, an abstraction map $\abstractionmap$, and a refinement map $\refinement$ such that
\begin{equation}
    \auxprobmeas^\numsamples \left( \probmeas^{\hat{\mc{M}},\scheduler}_{\initialsource}(\specification) \leq \probmeas^{\mc{M},\controller}_{\initialstate}(\specification), \;\forall\,\scheduler \in \schedulerset_{\hat{\mc{M}}} \right) \geq 1 - \confidenceparam,
\end{equation}
with the concrete policy $\controller = \refinement(\scheduler)$ and $\initialsource = \abstractionmap(\initialstate)$.
\end{problem}

The guarantee in \Cref{prob:main_data-driven} holds simultaneously for all schedulers on a single event of probability at least $1 - \confidenceparam$.
It therefore also applies to a scheduler selected by optimizing the learned abstraction using the same dataset, without requiring an additional confidence allocation.
This yields the following certified policy synthesis problem.

\begin{problem}[Control Synthesis]\label{prob:control_synthesis}
Given the inputs of \Cref{prob:main_data-driven}, construct an abstract model $\hat{\mc{M}}$, an abstraction map $\abstractionmap$, a refinement map $\refinement$, and a scheduler
\begin{equation}
    \scheduler^\star \in \argmax_{\scheduler \in \schedulerset_{\hat{\mc{M}}}} \probmeas^{\hat{\mc{M}},\scheduler}_{\initialsource}(\specification), \qquad \initialsource = \abstractionmap(\initialstate),
\end{equation}
such that
\begin{equation}
    \auxprobmeas^\numsamples \left(\probmeas^{\hat{\mc{M}},\scheduler^\star}_{\initialsource}(\specification) \leq \probmeas^{\mc{M},\refinement(\scheduler^\star)}_{\initialstate}(\specification)\right) \geq 1 - \confidenceparam.
\end{equation}
\end{problem}

\section{Data-Driven Abstraction Framework}
\label{sec:framework}

\Cref{prob:main_data-driven} is particularly challenging: both the system and the specification are complex, yet we do not have direct access to the system model. Instead, we must rely on sampled data. This problem has previously been considered with principled approaches to solving it under varying system assumptions (linear \cite{badings2023robust}, non-linear with additive noise \cite{nazeri2025data}, general non-linear \cite{nazeri2025data2}). The common feature for these methods is that the abstraction target model has been \glspl{imdp}. In contrast, as described in \Cref{sec:prelim}, recent research on model-based abstractions shows that factored representations are superior in terms of memory and frequently also accuracy \cite{mathiesen2025scalable}. Additionally, the abstraction process for \glspl{fimdp} remains similar to \gls{imdp}-based abstractions. Hence, intuitively, adapting data-driven techniques to \glspl{fimdp} requires only simple modifications. 
In \Cref{subsec:abstraction_process}, we describe the structure of the \glspl{fimdp} abstraction with an emphasis on the differences to the \gls{imdp}-based approaches, while in \Cref{subsec:learning}, we describe the data-driven framework to build the abstraction.

\subsection{Abstraction}\label{subsec:abstraction_process}
We describe below the general process of building the \gls{fimdp}-abstraction. The procedure is also summarized in Algorithm~\ref{alg:abstraction_process}.

\begin{algorithm}[t]
\caption{Data-driven \gls{fimdp} abstraction construction (summary of \Cref{subsec:abstraction_process})}
\label{alg:abstraction_process}
\begin{algorithmic}[1]
\Require State space $\statespace$, control space $\controlspace$, labeling $\labelling$, dataset $\dataset$, confidence parameter $\confidenceparam$.
\Ensure Abstract \gls{fimdp} $\hat{\mc{M}} = (\stateset, \actionset, h, \{\ambiguityset_{\source,\action}\}_{\source \in \stateset, \action \in \actionset})$.
\For{$i \gets 1$ to $n$}
    \State Partition $\statespace_i$ into regions $\{\region^i_1,\ldots,\region^i_{\numregions_i}\}$ and add complement region $\region^i_\star$.
    \State Define abstract marginal state set $\stateset_i = \{\source^i_1, \source^i_2, \ldots, \source^i_{\numregions_i}, \source^i_\star\}$.
\EndFor
\State Define joint state set $\stateset = \prod_{i=1}^n \stateset_i$ and abstraction map $\abstractionmap$.
\State Construct abstract labeling $h$ using the \emph{all}/\emph{no} convention over each $\abstractionmap^{-1}(\source)$.
\State Partition $\controlspace$ into regions $\{\auxregion_1,\ldots,\auxregion_{\numauxregions}\}$, define actions $\actionset$ and representatives $\overline{\control}_\action$.
\For{each $(\source,\action) \in \stateset \times \actionset$}
    \State Learn marginal ambiguity sets $\{\ambiguityset^i_{\source,\action}\}_{i=1}^n$ using \Cref{alg:learning}.
    \State Set joint ambiguity set $\ambiguityset_{\source,\action}$ as in Definition~\ref{def:fimdp}.
\EndFor
\State \Return $\hat{\mc{M}}$
\end{algorithmic}
\end{algorithm}

\textbf{States.} We partition the (hyperrectangular) state space $\statespace$ into an axis-aligned grid. That is, for each axis $i$, partition the compact set $\statespace_i$ into $\numregions_i \in \naturals$ disjoint Borel cells $\{\region^i_1, \region^i_2, \ldots, \region^i_{\numregions_i}\}$ such that the partition covers $\statespace_i$, i.e., $\bigcup_{j=1}^{\numregions_i} \region^i_j = \statespace_i$. The simplest partition of this type is one in which every region $\region_j^i$ is a half-open interval with the open side to the right except the right-most region, which is a fully closed interval.
This convention assigns each boundary point to exactly one cell, ensuring that the abstraction map is single-valued even when the successor distribution assigns positive probability to cell boundaries.

We also define a \emph{per-axis} complement region $\region^i_\star = \reals \setminus \statespace_i$ such that $\bigcup_{j=1}^{\numregions_i} \region^i_j \cup \region^i_\star$ covers $\reals$. In the abstract model, we associate each axis with a state variable whose values correspond to marginal regions; i.e., $\stateset_i = \{\source^i_1, \source^i_2, \ldots, \source^i_{\numregions_i}, \source^i_\star\}$. The joint abstract state space is defined as the product of the marginal state spaces $\stateset = \prod_{i=1}^n \stateset^i$.
To track the assignment, we define an abstraction map $\abstractionmap : \reals^n \to \stateset$ such that $\sysstate \in \region_\source$ if and only if $\abstractionmap(\sysstate) = \source$, and its preimage $\abstractionmap^{-1} : \stateset \to 2^{\reals^n}$. The set of initial abstract states $\initialstateset$ are the states $\source$ where there exists an $\sysstate \in \abstractionmap^{-1}(\source)$ such that $\sysstate \in \initialset$.

\begin{remark}[Memory Complexity]\label{rem:memory}
    Assume each axis is partitioned into $\kappa$ cells, so that $\lvert \stateset_i \rvert = \kappa$ and $\lvert \stateset \rvert = \kappa^n$.
    An \gls{imdp} stores a lower and an upper bound per source-action-destination triple, requiring
    \begin{equation}\label{eq:memory_imdp}
        2 \lvert \stateset \rvert^2 \lvert \actionset \rvert
        = 2\kappa^{2n} \lvert \actionset \rvert
    \end{equation}
    numbers. An \gls{fimdp} stores stores a lower and an upper bound for each marginal instead, requiring
    \begin{equation}\label{eq:memory_fimdp}
        2 \lvert \stateset \rvert \lvert \actionset \rvert \sum_{i=1}^n \lvert \stateset_i \rvert
        = 2n\kappa^{n+1} \lvert \actionset \rvert
    \end{equation}
    numbers. This reduces memory usage by a factor of $\kappa^{n-1}/n$, which grows exponentially in $n$ for any choice of $\kappa$.
    The joint ambiguity set $\ambiguityset_{\source,\action}$ need not be materialized since robust value iteration operates directly on the marginals~\cite{mathiesen2025scalable}.
\end{remark}

\textbf{Labelling.}
As each abstract state corresponds to a \emph{set} of concrete states, we must distinguish between whether \emph{all} or \emph{no} concrete states $\sysstate \in \abstractionmap^{-1}(\source)$ for an abstract are labelled with some label $\outputvar \in \outputspace$ \cite{cauchi2019efficiency}. 
Hence, in the abstract space, the output space $\hat{\outputspace}$ is augmented as $\hat{\outputspace} = \outputspace \cup \overline{\outputspace}$ where $\outputvar \in \outputspace$ is emitted for a state $\source$ if $\outputvar \in \labelling(\sysstate)$ for \emph{all} $\sysstate \in \abstractionmap^{-1}(\source)$ and $\overline{\outputvar} \in \overline{\outputspace}$ if $\outputvar \notin \labelling(\sysstate)$ for \emph{all} $\sysstate \in \abstractionmap^{-1}(\source)$.
Following~\cite{cauchi2019efficiency,9144391}, we write the specification in positive normal form and replace each $\neg\outputvar$ by $\overline{\outputvar}$ when evaluating it on the abstract model.
With slight abuse of notation, we re-use the symbol $\labelling : \stateset \to 2^{\hat{\outputspace}}$ for the labelling of the abstract model. 

\textbf{Actions.}
Similarly to the state space, we partition the control space $\controlspace$ into hyperrectangles $\{\auxregion_1, \auxregion_2, \ldots, \auxregion_\numauxregions\}$ and associate each with one abstract action, i.e., $\actionmap : \controlspace \to \actionset$. Furthermore, we select the center of each region $\overline{\control}_j$ as its representative input. That is, when deploying or refining the controller from the abstract model to the concrete system, if an action $\action_j \in \actionset$ is selected in the abstract model, then $\overline{\control}_j$ is emitted to the concrete system. 

\textbf{Transitions.}
For the abstraction to be useful, every behaviour of the concrete system must be reproducible by the abstract model. Otherwise, guarantees computed on the abstraction need not transfer to the system. We therefore design the abstract ambiguity sets $\ambiguityset_{\source,\action}$ such that every concrete transition can be matched by an equivalent abstract transition. Formally, fix any abstract state $\source \in \stateset$ and action $\action$. We require that for every $\sysstate \in \region_{\source}$ there exists a $\probdist_{\source,\action} \in \ambiguityset_{\source,\action}$ such that
\begin{equation}\label{eq:matching}
    \transitionkernel(\region_{\destination} \mid \sysstate, \overline{\control}_\action) = \probdist_{\source,\action}(\destination) \qquad \text{for all } \destination \in \stateset.
\end{equation}
Together with a few additional requirements, condition~\eqref{eq:matching} essentially induces an (alternating probabilistic) simulation relation between the concrete system and the abstract model, allowing guarantees to be transferred from the latter to the former.

\begin{definition}[Probabilistic Simulation Relation]
    A binary relation $\relation \subseteq \reals^n \times \stateset$ is a probabilistic simulation relation from $\hat{\mc{M}}$ to $\mc{M}$ if
    \begin{itemize}
        \item for every $\initialstate \in \initialset$, there exists a $\initialsource \in \initialstateset$ such that $(\initialstate, \initialsource) \in \relation$,
        \item for every $\source \in \stateset$, $\outputvar \in h(\source)$ if and only if $\outputvar \in h(\sysstate')$ for all $\sysstate' \in \{\sysstate \in \statespace : \sysstate \relation \source\}$, and similarly $\overline{\outputvar} \in h(\source)$ if and only if $\outputvar \notin h(\sysstate')$ for all $\sysstate' \in \{\sysstate \in \statespace : \sysstate \relation \source\}$, and
        \item for all $(\sysstate, \source)\in \relation$ we have $\forall \action \in \actionset$, $\exists \overline{\control}_\action \in \controlspace$ there exists a distribution $\probdist_{\source,\action} \in \ambiguityset_{\source,\action}$ with a coupling measure $\coupling$ on the measurable space $(\statespace \times \stateset, \borel(\statespace \times \stateset))$ such that
        \begin{itemize}
            \item $\coupling(\reals^n, \destination) = \probdist_{\source,\action}(\destination)$ for any $\destination \in \stateset$,
            \item $\coupling(B, \stateset) = \transitionkernel(B \mid \sysstate, \overline{\control}_\action)$ for any Borel measurable set $B \subseteq \reals^n$, and
            \item $\coupling(\relation) = 1$.
        \end{itemize}
    \end{itemize}
\end{definition}
If there exists a probabilistic simulation relation from $\hat{\mc{M}}$ to $\mc{M}$, we write $\hat{\mc{M}} \simrelation \mc{M}$. The idea of the simulation relation is to guarantee that the (probabilistic) behaviour of concrete system is captured by the abstract model, so that the verification result of the abstract model can be lifted to the concrete system. Moreover, notice how, for every pair $(\sysstate, \source) \in \relation$, each abstract action $\action$ can be matched by a control input $\overline{\control}_\action$ such that the concrete and abstract transition probabilities are related. The consequence is that \emph{any} scheduler $\scheduler$ can be refined to the concrete system and the verification result of the abstract model under this scheduler provides a lower bound on the satisfaction probability of the concrete system with the refined controller. Therefore, it is paramount to solve both \Cref{prob:main_data-driven} and \Cref{prob:control_synthesis} to prove that such a simulation relation exists for the constructed abstraction.

\begin{theorem}\label{thm:simulation_relation}
    Let $\hat{\mc{M}}$ be an abstraction of $\mc{M}$ constructed as described in \cref{subsec:abstraction_process}.
    If the ambiguity sets satisfy \eqref{eq:matching} for every abstract state $\source$ and action $\action$, then $\hat{\mc{M}} \simrelation \mc{M}$.
\end{theorem}
\begin{proof}
    We proceed with a proof by construction. In particular, we prove that the following relation is indeed a probabilistic simulation relation
    \begin{equation}
        \relation = \{ (\sysstate, \source) : \sysstate \in \region_\source \}.
    \end{equation}
    The initial-state condition follows from the definition
    $\initialstateset = \{ \source \in \stateset : \region_\source \cap \initialset \neq \emptyset \}$.
    The labelling condition follows directly from the construction.

    It remains to show that for each $(\sysstate, \source) \in \relation$ and for each $\action \in \actionset$ there exists a control action $\overline{\control}_\action \in \controlspace$, a distribution $\probdist_{\source, \action} \in \ambiguityset_{\source, \action}$, and a coupling probability measure $\coupling$ with the condition that $\coupling(\relation) = 1$. First, by \eqref{eq:matching}, we can pick the distribution $\probdist_{\source, \action}$ such that $\transitionkernel(\region_\destination \mid \sysstate, \overline{\control}_\action) = \probdist_{\source, \action}(\destination)$ for all $\destination \in \stateset$.
    Then, define the coupling measure as 
    \begin{equation}
        \coupling(B, \tilde{\stateset}) = \transitionkernel\left(B \cap \abstractionmap^{-1}(\tilde{\stateset})
        \;\middle|\;
        \sysstate,\overline{\control}_\action
    \right) 
    \end{equation}
    for any Borel measurable set $B \in \borel(\reals^n)$ and any set $\tilde{\stateset} \subseteq \stateset$.
    Clearly, $\coupling(B, \stateset) = \transitionkernel(B \mid \sysstate, \overline{\control}_\action)$ for any Borel measurable set $B \subseteq \reals^n$ and $\coupling(\reals^n, \destination) = \transitionkernel(\region_\destination \cap \reals^n \mid \sysstate, \overline{\control}_\action) = \probdist_{\source, \action}(\destination)$ for any $\destination \in \stateset$. Finally, the relation is contained with probability one, $\coupling(\relation) = \sum_{\destination \in \stateset} \coupling(\region_\destination, \destination) = 1$, since $\bigcup_{\destination \in \stateset} \region_\destination$ is a partition of $\reals^n$, which concludes the proof. 
\end{proof}

\Cref{thm:simulation_relation} ensures that any scheduler $\scheduler$ can be refined to the concrete system.
The procedure for refining $\scheduler$ is as follows: for a given state $\sysstate$ and \gls{dpa} state $\automatonstate$ (with the \gls{dpa} internal to $\scheduler$), the control input to the concrete system is $\overline{\control}_{\action}$ where $\action = \scheduler_\automatonstate(\source)$, $\source = \abstractionmap(\sysstate)$ and the scheduler transitions to $\scheduler_{\automatonstate^+}$ where $\automatonstate^+ = \automatontransition(\automatonstate, \labelling(\sysstate^{+}))$. This construction is the refinement map $\refinement$. The satisfaction probability starting from $\sysstate$ and $\automatonstate$ (with the system controlled by $\refinement(\scheduler)$) is lower-bounded by the satisfaction probability for $\source$ and $\automatonstate$. 

\begin{remark}
    Note that if any $f_i$, $i \in \{1, \ldots, n \}$ depends only on a subset of $\staterandomvar[\timevar]$, then one can exploit the sparsity to achieve greater scalability and sample complexity \cite{SAM2015DBN,SAM2017DBN}. We defer exploring this sparsity to future work.
\end{remark}

The challenge in building the abstraction is to construct the ambiguity sets to satisfy~\eqref{eq:matching}; the following subsection focuses on addressing this.

\subsection{Learning the marginal ambiguity sets from data}\label{subsec:learning}
We do not have access to $f$. Instead, we rely on a dataset $\dataset = \{\sample_1, \ldots, \sample_\numsamples\}$ of noisy state-action-state observations $\sample_j = (\sysstate_j, \control_j, \sysstate_j^+)$ with $\sysstate_j^+ = f(\sysstate_j, \control_j, \noisebasic_j)$. The noise realizations $\noisebasic_j$ are i.i.d., independent of the state-action pairs $(\sysstate_j, \control_j)$, and are not recorded in $\dataset$. For each source-action pair $(\source, \action)$, the marginal ambiguity sets are constructed in five steps described below; the steps are also summarized in Algorithm~\ref{alg:learning}.

\begin{algorithm}[t]
\caption{Learning marginal ambiguity sets from data (for one $(\source,\action)$)}
\label{alg:learning}
\begin{algorithmic}[1]
\Require Dataset $\dataset = \{(\sysstate_j,\control_j,\sysstate^+_j)\}_{j=1}^{\numsamples}$, source region $\region_\source$, action region $\auxregion_\action$ with representative input $\overline{\control}_\action$, confidence parameter $\confidenceparam$.
\Ensure Marginal ambiguity sets $\{\ambiguityset^i_{\source,\action}\}_{i=1}^n$.

\State $\overline{\confidenceparam} \gets \confidenceparam / (2\lvert \stateset \rvert \lvert \actionset \rvert \sum_{i=1}^n \lvert \stateset_i \rvert)$ \Comment{Per-transition confidence budget.}

\Statex \textit{(1) Split the dataset:}
\State $\dataset_{\source,\action} \gets \{(\sysstate_j,\control_j,\sysstate^+_j)\in\dataset: \sysstate_j\in\region_\source,\ \control_j\in\auxregion_\action\}$
\State $\numsamples_{\source,\action} \gets |\dataset_{\source,\action}|$

\Statex \textit{(2--3) Over-approx. reachable sets and project to marginals:}
\For{each sample $(\sysstate_j,\control_j,\sysstate^+_j) \in \dataset_{\source,\action}$}
    \For{$i \gets 1$ to $n$}
        \State $\mytolerance^i_j \gets \lipschitzconst^i_\statespace\sup_{\hat{\sysstate}\in\region_\source}\|\sysstate_j-\hat{\sysstate}\| + \lipschitzconst^i_\controlspace\|\control_j-\overline{\control}_\action\|$
        \State $\hat{\frs}^i_j \gets [\sysstate^+_{j,i}-\mytolerance^i_j,\ \sysstate^+_{j,i}+\mytolerance^i_j]$
    \EndFor
\EndFor

\Statex \textit{(4) Count marginal transitions:}
\For{$i \gets 1$ to $n$}
    \For{each $\destination_i \in \stateset_i$}
        \State $\underline{\numsamples}^i_{\source,\action}(\destination_i) \gets |\{j: \hat{\frs}^i_j \subseteq \region_{\destination_i}\}|$
        \State $\overline{\numsamples}^i_{\source,\action}(\destination_i) \gets |\{j: \hat{\frs}^i_j \cap \region_{\destination_i} \neq \emptyset\}|$
    \EndFor
\EndFor

\Statex \textit{(5) Clopper-Pearson bounds and ambiguity sets:}
\For{$i \gets 1$ to $n$}
    \For{each $\destination_i \in \stateset_i$}
        \State $\underline{\probdist}^i_{\source,\action}(\destination_i) \gets \mathrm{B}^{-1}(\overline{\confidenceparam};\ \underline{\numsamples}^i_{\source,\action}(\destination_i),\ \numsamples_{\source,\action}-\underline{\numsamples}^i_{\source,\action}(\destination_i)+1)$
        \State $\overline{\probdist}^i_{\source,\action}(\destination_i) \gets \mathrm{B}^{-1}(1-\overline{\confidenceparam};\ \overline{\numsamples}^i_{\source,\action}(\destination_i)+1,\ \numsamples_{\source,\action}-\overline{\numsamples}^i_{\source,\action}(\destination_i))$
    \EndFor
    \State $\ambiguityset^i_{\source,\action} \gets \{\probdist^i\in\dists(\stateset_i): \underline{\probdist}^i_{\source,\action}(\destination_i)\leq \probdist^i(\destination_i)\leq \overline{\probdist}^i_{\source,\action}(\destination^i),\ \forall\destination^i\in\stateset^l\}$
\EndFor
\State \Return $\{\ambiguityset^i_{\source,\action}\}_{i=1}^n$
\end{algorithmic}
\end{algorithm}

\begin{figure}
    \centering
    \includegraphics[width=0.7\linewidth]{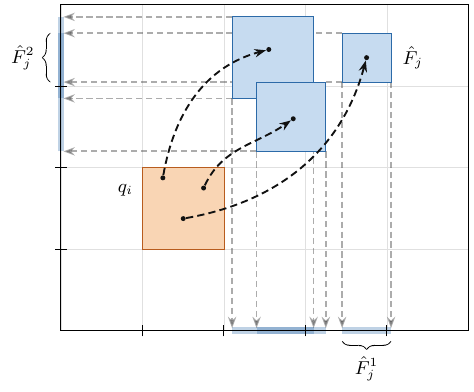}
    \caption{Depiction of steps 2-4 of the ambiguity set learning process. Samples that start from $\region_i$ are expanded to cover the entire $\region_i$ using known Lipschitz constants. The resulting over-approximations of the forward reachable sets are projected onto each axis/marginal where they are counted if they overlap or completely contain an abstract (marginal) state.}
    \Description{Three samples of the form $(x_j, u_j, x_j^+)$ with $x_j$ in region $q_i$ (orange) are depicted along with $x_j^+$ ($u_j$ is not depicted). A hyperrectangle $\hat{F}_j$ (blue) is expanded around $x_j^+$ according to known Lipschitz constants (see Lemma~\ref{lem:overapprox}) such that any state $\tilde{x}$ in $q_i$ will land in $\hat{F}_j$ under the system dynamics $f$, the control $u_j$, and the unknown noise $v_j$. Each $\hat{F}_j$ is projected onto the axes/marginals; each marginal corresponds to a state variable and the abstract marginal states are depicted as green balls in the segment they abstractly represent.}
    \label{fig:learning_process_pictorial}
\end{figure}

\paragraph{1) Splitting the dataset.}
We collect the samples whose state and input lie in the regions associated with $\source$ and $\action$,
\begin{equation}
    \dataset_{\source,\action} = \big\{ (\sysstate_j, \control_j, \sysstate^+_j) \in \dataset : \sysstate_j \in \region_\source,\ \control_j \in \auxregion_\action \big\},
\end{equation}
and write $\numsamples_{\source,\action} = \lvert \dataset_{\source,\action} \rvert$.

\paragraph{2) Forward reachable sets.}
For a noise value $\noisebasic$, the forward reachable set of the region $\region_\source$ under the representative input $\overline{\control}_\action$ is
\begin{equation}
    \frs_{\source,\action}(\noisebasic) \coloneqq \left\{ f(\sysstate, \overline{\control}_\action, \noisebasic) : \sysstate \in \region_\source \right\}.
\end{equation}
This set contains the successor of \emph{every} state in $\region_\source$ under the same noise realization. Since neither $f$ nor the noise realizations are known, we over-approximate it using the samples. For each sample $(\sysstate_j, \control_j, \sysstate^+_j) \in \dataset_{\source,\action}$, define the box $\hat{\frs}_j = \prod_{i=1}^n \hat{F}^i_j$ with
\begin{equation}
    \hat{\frs}^i_j \coloneqq \left[ \sysstate^+_{j,i} - \mytolerance^i_j,\ \sysstate^+_{j,i} + \mytolerance^i_j \right], \quad
    \mytolerance^i_j \coloneqq \lipschitzconst^i_\statespace \sup_{\hat{\sysstate} \in \region_\source} \|\sysstate_j - \hat{\sysstate}\| + \lipschitzconst^i_\controlspace \|\control_j - \overline{\control}_\action\|.
\end{equation}

\begin{lemma}[Data-driven over-approximation]\label{lem:overapprox}
    For every sample $(\sysstate_j, \control_j, \sysstate^+_j) \in \dataset_{\source,\action}$, it holds that $\frs_{\source,\action}(\noisebasic_j) \subseteq \hat{\frs}_j$.
\end{lemma}

\begin{proof}
    Let $y \in \frs_{\source,\action}(\noisebasic_j)$, i.e., $y = f(\sysstate, \overline{\control}_\action, \noisebasic_j)$ for some $\sysstate \in \region_\source$. Since $y$ and $\sysstate^+_j$ are generated by the \emph{same} noise realization $\noisebasic_j$, \eqref{eq:lipschitz} gives, for each $i$,
    \begin{equation}
        \begin{aligned}
        |y_i - \sysstate^+_{j,i}| &= |f_i(\sysstate, \overline{\control}_\action, \noisebasic_j) - f_i(\sysstate_j, \control_j, \noisebasic_j)| \\
        &\leq \lipschitzconst^i_\statespace \|\sysstate - \sysstate_j\| + \lipschitzconst^i_\controlspace \|\overline{\control}_\action - \control_j\|
        \leq \mytolerance^i_j,
        \end{aligned}
    \end{equation}
    where the last step uses $\sysstate \in \region_\source$. Hence $y_i \in \hat{\frs}^i_j$ for all $i \in \{1, \ldots, n\}$.
\end{proof}

\paragraph{3) Projection onto the marginals.}
Rather than reasoning about the $n$-dimensional sets $\frs_{\source,\action}(\noisebasic)$ and $\hat{\frs}_j$, we only use their projections onto each axis,
\begin{equation}
    \frs^i_{\source,\action}(\noisebasic) \coloneqq \projection_i\left(\frs_{\source,\action}(\noisebasic)\right) = \left\{ f_i(\sysstate, \overline{\control}_\action, \noisebasic) : \sysstate \in \region_\source \right\}
\end{equation}
and $\projection_i(\hat{\frs}_j) = \hat{\frs}^i_j$.
Since projection preserves inclusion, Lemma~\ref{lem:overapprox} implies $\frs^i_{\source,\action}(\noisebasic_j) \subseteq \hat{\frs}^i_j$ for all $i$ and $j$. The projected reachable sets bound the marginals of the transition kernel uniformly over the source region.

\begin{lemma}[Uniform marginal sandwich]\label{lem:sandwich}
    For all $i \in \{1, \ldots, n\}$, $\destination_i \in \stateset_i$ and all $\sysstate \in \region_\source$,
    \begin{equation}\label{eq:sandwich}
        \underbrace{\probmeas\left[ \frs^i_{\source,\action}(\noisebasic) \subseteq \region_{\destination_i} \right]}_{\underline{\theta}^i_{\source,\action}(\destination_i)}
        \;\leq\; \transitionkernel_i(\region_{\destination_i} \mid \sysstate, \overline{\control}_\action)
        \;\leq\; \underbrace{\probmeas\left[ \frs^i_{\source,\action}(\noisebasic) \cap \region_{\destination_i} \neq \emptyset \right]}_{\overline{\theta}^i_{\source,\action}(\destination_i)}.
    \end{equation}
\end{lemma}

\begin{proof}
    Fix $\sysstate \in \region_\source$. By definition, we have that $\transitionkernel_i(\region_{\destination_i} \mid \sysstate, \overline{\control}_\action) = \probmeas\left[ f_i(\sysstate, \overline{\control}_\action, \noisebasic) \in \region_{\destination_i} \right]$. Since $f_i(\sysstate, \overline{\control}_\action, \noisebasic) \in \frs^i_{\source,\action}(\noisebasic)$ for every $\noisebasic$, it holds
    \begin{equation}
        \left\{ \frs^i_{\source,\action}(\noisebasic) \subseteq \region_{\destination_i} \right\}
        \subseteq \left\{ f_i(\sysstate, \overline{\control}_\action, \noisebasic) \in \region_{\destination_i} \right\}
        \subseteq \left\{ \frs^i_{\source,\action}(\noisebasic) \cap \region_{\destination_i} \neq \emptyset \right\},
    \end{equation}
    and monotonicity of probability yields~\eqref{eq:sandwich}. Note that $\underline{\theta}_{\source,\action}(\destination_i)$ and $\overline{\theta}_{\source,\action}(\destination_i)$ do not depend on $\sysstate$.
\end{proof}

This projection onto the marginals is the key difference to \gls{imdp}-based approaches~\cite{nazeri2025data}. Those approaches check, for each joint region $\region_\destination$, whether a box $\hat{\frs}_j$ is contained in or intersects it. A box only counts as contained if it lies inside $\region_\destination$ along \emph{all} axes simultaneously.  The marginal approach instead requires containment along one axis at a time and estimates only $\sum_{i=1}^n |\stateset_i|$ probabilities. This yields substantially fewer probabilities to estimate. The independent noise assumption (Assumption~\ref{ass:independent_noise}) guarantees that no soundness is lost by doing so.

\paragraph{4) Counting marginal transitions.}
For each $i$ and $\destination_i \in \stateset_l$, we count the projected reachable sets that are contained in and intersect the interval $\region_{\destination^l}$:
\begin{subequations}
\begin{align}
    \underline{\numsamples}^i_{\source,\action}(\destination_i) &= \left|\{ j : \hat{\frs}^i_j \subseteq \region_{\destination_i} \}\right|, \\
    \overline{\numsamples}^i_{\source,\action}(\destination_i) &= \left|\{ j : \hat{\frs}^i_j \cap \region_{\destination_i} \neq \emptyset \}\right|.
\end{align}
\end{subequations}

\paragraph{5) Clopper-Pearson bounds.}
Let $\confidenceparam \in (0,1)$ and let $\mathrm{B}^{-1}(\theta; a, b)$ denote the $\theta$-quantile of the Beta distribution with parameters $a, b$. We set
\begin{subequations}
\begin{align}
    \underline{\probdist}^i_{\source,\action}(\destination_i) &= \mathrm{B}^{-1}\left(\overline{\confidenceparam};\ \underline{\numsamples}^i_{\source,\action}(\destination_i),\ \numsamples_{\source,\action} - \underline{\numsamples}^i_{\source,\action}(\destination_i) + 1\right), \\
    \overline{\probdist}^i_{\source,\action}(\destination_i) &= \mathrm{B}^{-1}\left(1 - \overline{\confidenceparam};\ \overline{\numsamples}^i_{\source,\action}(\destination_i) + 1,\ \numsamples_{\source,\action} - \overline{\numsamples}^i_{\source,\action}(\destination_i)\right),
\end{align}
\end{subequations}
with the conventions $\underline{\probdist}^i_{\source,\action}(\destination_i) = 0$ if $\underline{\numsamples}^i_{\source,\action}(\destination_i) = 0$ and $\overline{\probdist}^i_{\source,\action}(\destination_i) = 1$ if $\overline{\numsamples}^i_{\source,\action}(\destination_i) = N$ and $\overline{\confidenceparam} = \confidenceparam / (2\lvert \stateset \rvert \lvert \actionset \rvert \sum_{i=1}^n \lvert \stateset_i \rvert)$. The marginal ambiguity set is the interval set
\begin{equation}
    \ambiguityset^i_{\source,\action} = \left\{ \probdist^i_{\source,\action} \in \dists(\stateset^i) : \underline{\probdist}^i_{\source,\action}(\destination_i) \leq \probdist^i_{\source,\action}(\destination_i) \leq \overline{\probdist}^i_{\source,\action}(\destination_i), \, \forall \destination_i \in \stateset_i \right\},
\end{equation}
and $\ambiguityset_{\source,\action}$ is defined as in Definition~\ref{def:fimdp}.

\paragraph{Soundness.}
The following theorem shows that the ambiguity sets constructed in steps~1-5 satisfy the matching condition~\eqref{eq:matching} with high confidence.

\begin{theorem}[Soundness]\label{thm:soundness}
    Fix $(\source, \action)$ and assume the dynamics are of the form in Assumption~\ref{ass:independent_noise}. Then, given a dataset $\dataset$ sampled according to $\auxprobmeas^\numsamples$, with probability at least $1 - \confidenceparam / \lvert \stateset \rvert \lvert \actionset \rvert$ over $\auxprobmeas^\numsamples$, the following holds: for all $\sysstate \in \region_\source$, there exists a distribution $\probdist_{\source,\action} \in \ambiguityset_{\source,\action}$ such that $\probdist_{\source,\action}(\destination) = \prod_{i=1}^n \transitionkernel_i(\region_{\destination_i} \mid \sysstate, \overline{\control}_\action)$ for all $\destination \in \stateset$.
\end{theorem}
\begin{proof}
    Conditioning on the state-action pairs fixes $\dataset_{\source,\action}$, its size $\numsamples_{\source,\action}$, and the radii $\mytolerance^i_j$. Since the noise realizations are \gls{iid} and independent of the state-action pairs, the $\noisebasic_j$ indexing $\dataset_{\source,\action}$ remain \gls{iid} after conditioning. Fix $i$ and $\destination_i \in \stateset_i$.

    \emph{Counts for the true reachable sets.}
    Define the (unobservable) counts
    \begin{subequations}
    \begin{align}
        \underline{M}^i_{\source,\action}(\destination_i) &= \left|\{ j : \frs^i_{\source,\action}(\noisebasic_j) \subseteq \region_{\destination_i} \}\right|, \\
        \overline{M}^i_{\source,\action}(\destination_i) &= \left|\{ j : \frs^i_{\source,\action}(\noisebasic_j) \cap \region_{\destination_i} \neq \emptyset \}\right|.
    \end{align}
    \end{subequations}
    The set $\frs^i_{\source,\action}(\noisebasic_j)$ depends on sample $j$ only through $\noisebasic_j$. Hence, each count is a sum of $N$ \gls{iid} Bernoulli indicators, with $\underline{M}^i_{\source,\action}(\destination_i) \sim \mathrm{Bin}(\numsamples_{\source,\action}, \underline{\theta}^i_{\source,\action}(\destination_i))$ and $\overline{M}^i_{\source,\action}(\destination_i) \sim \mathrm{Bin}(\numsamples_{\source,\action}, \overline{\theta}^i_{\source,\action}(\destination_i))$. By the one-sided Clopper-Pearson guarantees, the events
    \begin{subequations}\label{eq:cp-true}
    \begin{align}
        \underline{\mathcal{E}}^i_{\source,\action}(\destination_i) &= \left\{ \mathrm{B}^{-1}\left(\overline{\confidenceparam};\, \underline{M}^i_{\source,\action}(\destination_i),\, \numsamples_{\source,\action} - \underline{M}^i_{\source,\action}(\destination_i) + 1\right) \leq \underline{\theta}^i_{\source,\action}(\destination_i) \right\}, \label{eq:cp-true-lower} \\
        \overline{\mathcal{E}}^i_{\source,\action}(\destination_i) &= \left\{ \overline{\theta}^i_{\source,\action}(\destination_i) \leq \mathrm{B}^{-1}\left(1 - \overline{\confidenceparam};\, \overline{M}^i_{\source,\action}(\destination_i) + 1,\, \numsamples_{\source,\action} - \overline{M}^i_{\source,\action}(\destination_i)\right) \right\} \label{eq:cp-true-upper}
    \end{align}
    \end{subequations}
    each have probability at least $1 - \overline{\confidenceparam} = 1 - \confidenceparam/(2\lvert \stateset \rvert \lvert \actionset \rvert \sum_{i=1}^n \lvert \stateset_i \rvert)$, with the boundary conventions of step~5.

    \emph{The data-driven counts are conservative.}
    By step~3, $\frs^i_{\source,\action}(\noisebasic_j) \subseteq \hat{\frs}^i_j$. Thus, $\hat{\frs}^i_j \subseteq \region_{\destination_i}$ implies $\frs^i_{\source,\action}(\noisebasic_j) \subseteq \region_{\destination_i}$, and $\frs^i_{\source,\action}(\noisebasic_j) \cap \region_{\destination_i} \neq \emptyset$ implies $\hat{\frs}^i_j \cap \region_{\destination_i} \neq \emptyset$. Hence, for every noise realization, $\underline{\numsamples}^i_{\source,\action}(\destination_i) \leq \underline{M}^i_{\source,\action}(\destination_i)$ and $\overline{M}^i_{\source,\action}(\destination_i) \leq \overline{\numsamples}^i_{\source,\action}(\destination_i)$. Both Clopper-Pearson endpoints are non-decreasing in the count. Therefore, on $\underline{\mathcal{E}}^i_{\source,\action}(\destination_i) \cap \overline{\mathcal{E}}^i_{\source,\action}(\destination_i)$, we have $\underline{\probdist}^i_{\source,\action}(\destination_i) \leq \underline{\theta}^i_{\source,\action}(\destination_i)$ and $\overline{\theta}^i_{\source,\action}(\destination_i) \leq \overline{\probdist}^i_{\source,\action}(\destination_i)$.

    \emph{Uniformity over $\region_\source$.}
    Combining this with Lemma~\ref{lem:sandwich}, on $\underline{\mathcal{E}}^i_{\source,\action}(\destination_i) \cap \overline{\mathcal{E}}^i_{\source,\action}(\destination_i)$, it holds that
    \begin{equation}
        \underline{\probdist}^i_{\source,\action}(\destination_i) \leq \underline{\theta}^i_{\source,\action}(\destination_i) \leq \transitionkernel_i(\region_{\destination_i} \mid \sysstate, \overline{\control}_\action) \leq \overline{\theta}^i_{\source,\action}(\destination_i) \leq \overline{\probdist}^i_{\source,\action}(\destination_i),
        \quad \forall \sysstate \in \region_\source.
    \end{equation}
    Neither the events nor $\underline{\theta}^i_{\source,\action}(\destination_i)$ and $\overline{\theta}^i_{\source,\action}(\destination_i)$ depend on $\sysstate$, so this is true simultaneously for all $\sysstate \in \region_\source$.

    \emph{Union bound and decomposition.}
    Let 
    \begin{equation}
    \mathcal{E}_{\source,\action} = \bigcap_{i=1}^n \bigcap_{\destination_i \in \stateset_i} \left( \underline{\mathcal{E}}^i_{\source,\action}(\destination_i) \cap \overline{\mathcal{E}}^i_{\source,\action}(\destination_i) \right).
    \end{equation}
    By the union bound, $\probmeas[\mathcal{E}_{\source,\action}] \geq 1 - \overline{\confidenceparam}\cdot(2\sum_{i=1}^n \lvert \stateset_i \rvert) = 1 - \confidenceparam/(\lvert \stateset \rvert \lvert \actionset \rvert)$. On $\mathcal{E}_{\source,\action}$, for every $\sysstate \in \region_\source$ and every $i$, the marginal $\left(\transitionkernel_i(\region_{\destination_i} \mid \sysstate, \overline{\control}_\action)\right)_{\destination_i \in \stateset_i}$ is a probability distribution over $\stateset_i$ that belongs by construction to $\ambiguityset^i_{\source,\action}$. By~\eqref{eq:product-ambiguity}, $\probdist_{\source,\action}(\destination) = \prod_{i=1}^n \transitionkernel_i(\region_\destination \mid \sysstate, \overline{\control}_\action) \in \ambiguityset_{\source,\action}$ for all $\destination \in \stateset$. Finally, Assumption~\ref{ass:independent_noise} gives, for all $\destination \in \stateset$,
    \begin{equation}
        \transitionkernel(\region_{\destination} \mid \sysstate, \overline{\control}_\action) = \prod_{i=1}^n \transitionkernel_i(\region_{\destination_i} \mid \sysstate, \overline{\control}_\action) = \probdist_{\source,\action}(\destination),
    \end{equation}
    which is~\eqref{eq:matching}.
\end{proof}
Note that Theorem~\ref{thm:soundness} is equivalent to an ambiguity set $\ambiguityset_{\source,\action}$ satisfying the requirement in~\eqref{eq:matching} with high probability for a given transition kernel and state-action pair $(\source,\action)$.

\begin{remark}[Sample Complexity]\label{rem:per_transition_confidence}
    The \gls{fimdp} has a factor $\kappa^{n-1}/n$ fewer transitions than the \gls{imdp} (see \Cref{rem:memory}).
    Under the union bound, it can therefore afford a larger per-transition confidence parameter:
    \begin{equation*}
        \overline{\confidenceparam}_{\mathrm{I}} = \frac{\confidenceparam}{2\kappa^{2n}\lvert \actionset \rvert},
        \qquad
        \overline{\confidenceparam}_{\mathrm{F}} = \frac{\confidenceparam}{2n\kappa^{n+1}\lvert \actionset \rvert}
        = \frac{\kappa^{n-1}}{n}\,\overline{\confidenceparam}_{\mathrm{I}}.
    \end{equation*}
    That is, the confidence \emph{budget} per transition is larger for \glspl{fimdp}.
    By Hoeffding's inequality, each one-sided bound of a transition interval of half-width $\varepsilon$
    fails with probability at most $\overline{\confidenceparam}$ for every sample size
    \begin{equation*}
        \numsamples_{\source,\action} \geq \underline{\numsamples}(\overline{\confidenceparam}, \varepsilon)
        \coloneqq \frac{\ln(1/\overline{\confidenceparam})}{2\varepsilon^2}.
    \end{equation*}
    The threshold $\underline{\numsamples}$ is decreasing in $\overline{\confidenceparam}$ for a fixed $\varepsilon$.
    Moreover, $\overline{\confidenceparam}_{\mathrm{F}} \geq \overline{\confidenceparam}_{\mathrm{I}}$ holds since $\kappa^{n-1} \geq n$ for all $\kappa \geq 2$ and $n \geq 1$.
    Hence, every sample size that suffices for the \gls{imdp} also suffices for the \gls{fimdp}.
    For the same $\varepsilon$ and $\confidenceparam \in (0,1)$, the two thresholds compare as
    \begin{equation}
        \frac{\underline{\numsamples}(\overline{\confidenceparam}_{\mathrm{F}}, \varepsilon)}{\underline{\numsamples}(\overline{\confidenceparam}_{\mathrm{I}}, \varepsilon)}
        = \frac{\ln\!\big(2 n \kappa^{n+1} \lvert \actionset \rvert / \confidenceparam\big)}{\ln\!\big(2 \kappa^{2n} \lvert \actionset \rvert / \confidenceparam\big)} \leq 1,
    \end{equation}
    with strict inequality whenever $\kappa^{n-1} > n$; equality only holds for $n = 1$, or $n = 2$ and $\kappa = 2$.
    In particular, the smallest admissible sample sizes satisfy
    $\lceil \underline{\numsamples}(\overline{\confidenceparam}_{\mathrm{F}}, \varepsilon) \rceil
    \leq \lceil \underline{\numsamples}(\overline{\confidenceparam}_{\mathrm{I}}, \varepsilon) \rceil$.
\end{remark}

\begin{corollary}\label{cor:high_confidence_simulation_relation}
    Given a concrete system $\mc{M}$ with unknown dynamics that satisfies Assumption~\ref{ass:independent_noise}, a dataset $\dataset$ sampled according to $\auxprobmeas^\numsamples$, then build an abstraction $\hat{\mc{M}}$ according to Subsection~\ref{subsec:abstraction_process} with the ambiguity sets constructed as in Theorem~\ref{thm:soundness}. Then, with probability at least $1 - \confidenceparam$ over $\auxprobmeas^\numsamples$, it holds that $\hat{\mc{M}} \simrelation \mc{M}$. 
\end{corollary}
The proof of Corollary~\ref{cor:high_confidence_simulation_relation} is the application of \Cref{thm:soundness} to each state-action pair, which are subsequently combined using a union-bound argument. A consequence of the corollary is that verification and control synthesis can be performed over $\hat{\mc{M}}$, which can be lifted to $\mc{M}$ by the refinement map $\refinement$. 

Note that due to the data-driven abstraction, every upper bound is non-zero. That is, we cannot distinguish whether a transition does not exist or that a transition is simply not observed in the finite dataset. Consequently, the transition graph is (existentially) complete, which makes a certain subclass of $\omega$-regular objectives trivial. We analyse this subclass and propose an algorithm for checking the triviality a priori in Appendix~\ref{app:trivial}.

\begin{figure*}[t]
  \centering
  \includegraphics[width=\linewidth]{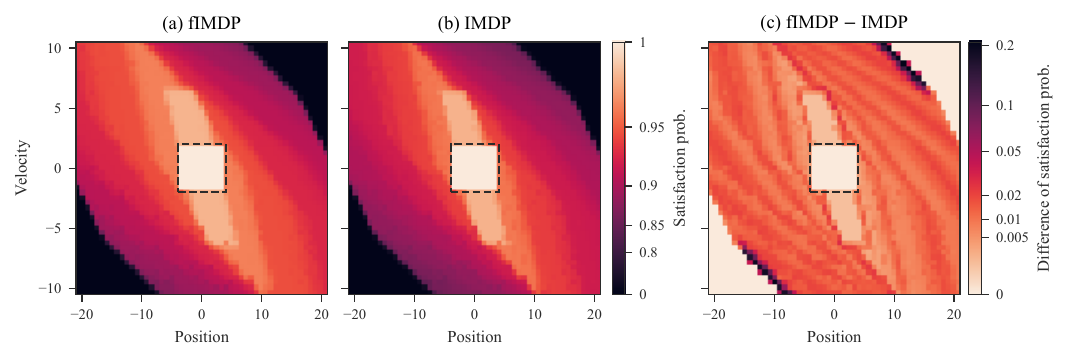}
  \caption{Satisfaction probabilities for the double integrator benchmark in the \gls{fimdp} and \gls{imdp} abstractions, and their difference on the right. The dashed box denotes the goal region. The \gls{fimdp} satisfaction probabilities are higher for all cells (up to $20$ percentage points). We used a power-law normalization $\left(\frac{v - v_{\min}}{v_{\max} - v_{\min}}\right)^\gamma$ of the satisfaction probabilities for better colour differentiation towards $1$. For panels (a) and (b), we use $v_{\min}=0$, $v_{\max}=1$, and $\gamma=8$, and for panel (c), we use $v_{\min}=0$, $v_{\max}$ as the maximum difference, and $\gamma=0.4$.}
  \Description{The figure includes three panels with the satisfaction probability for an fIMDP, an IMDP, and the difference in satisfaction probability between fIMDP and IMDP across the state space of the double integrator. The x-axis is the position and the y-axis is the velocity. The center contains a dashed box, which is the goal region. The non-zero lower bounds for both methods are in the shape of a rotated ellipse with higher probabilities towards the center. The difference in satisfaction probability is everywhere non-negative, i.e., in favour of the fIMDP.}
  \label{fig:satprob_comparison}
\end{figure*}

\section{Experimental Evaluation}
\label{sec:experiments}

We evaluate our proposed framework on benchmark case studies. For each case study, we report (i) the lower bound on the probability of satisfying the specification from the initial state, (ii) the computation time required to construct the abstraction from data, and (iii) the memory required to store the abstraction.
We choose to report only the construction time, as the synthesis/verification time depends on the solver employed, being aware that the different target models can impact the solving times as well. Similarly, we report the memory required to store the abstraction for the most comparable quantity, with it being independent of the runtime memory, garbage collection, etc. All implementations construct sparse models, hence, the memory usage of the model is also a proxy for the sparsity obtained by the respective construction methods.

To show the effectiveness of our \gls{fimdp} approach, we compare our results with the state-of-the-art in \cite{nazeri2025data2}, which addresses the same setting but without factorization. We implemented our experiments using Python with Jax~\cite{jax2018github} for efficient computation, similar to an implementation of~\cite{nazeri2025data2}, which we compare against. We run all experiments on a personal machine with an Apple M3 Pro CPU and $18$GB of RAM. 

In each experiment, the dynamics are only used to generate samples and are unknown to our framework. We use $[1000, 5000, 25000]$ samples per state-action pair, and we set the overall confidence to be at least $1-\beta = 0.95$. The results are reported in Tables~\ref{tab:memory} (memory) and \ref{tab:prob_time} (satisfaction probability and construction time).

\subsection{Double Integrator}
\textbf{System dynamics.} The double integrator describes a unit-mass moving along an axis. Its state is position and velocity: $x[k] = [p[k], v[k]]^\top$, and its input $u[k]$ is the acceleration. The discrete-time dynamics are
\begin{equation*}
    x[k+1] = \begin{bmatrix} 1 & \tau \\ 0 & 1 \end{bmatrix} x[k] + \begin{bmatrix} \tau^2/2 \\ \tau \end{bmatrix} u[k] + w[k],
\end{equation*}
where the components of $w[k]$ are i.i.d.\ with a triangular distribution supported on $[-0.15, 0.15]^2$ with the peak for either dimension at $0.0$. We use $\tau = 1\,\mathrm{s}$. The input set is the grid $\mathcal{U} = \{-5, -4.5, \dots, 4.5, 5\}$ with $21$ inputs.

\noindent\textbf{Abstraction.} The state space $\mathcal{X} = [-21, 21] \times [-10.5, 10.5]$ is partitioned into $51 \times 51 = 2{,}601$ regions. The system starts at $x[0] = [0, -8]^\top$ and must reach the goal set $\mathcal{G} = [-4, 4] \times [-2, 2]$.

\subsection{Pendulum}
\textbf{System dynamics.} The pendulum is a uniform rod of mass $m$ and length $l$ pivoted at one end. Its state is the angle from the upright position and the angular velocity: $x[k] = [\theta[k], \omega[k]]^\top$. The input $u[k]$ is a torque applied at the pivot, and $b$ is the damping coefficient. The discrete-time dynamics are
\begin{subequations}
\begin{align*}
    \theta[k+1] &= \operatorname{wrap}\left(\theta[k] + \tau\, \omega[k] + w_\theta[k]\right), \\
    \omega[k+1] &= (1 - b)\, \omega[k] + \tau \left( \frac{3g}{2l} \sin\theta[k] + \frac{3}{m l^2} u[k] \right) + w_\omega[k],
\end{align*}
\end{subequations}
where $w_\theta[k]$ and $w_\omega[k]$ are i.i.d.\ zero-mean Gaussian noise terms with standard deviations $\sigma_\theta$ and $\sigma_\omega$, and $\operatorname{wrap}(\theta) \triangleq ((\theta + \pi) \bmod 2\pi) - \pi$ maps the angle back to $[-\pi, \pi)$\footnote{$\operatorname{wrap}$ is not Lipschitz continuous. However, the discontinuity is merely an artefact of representing the angular space $\mathbb{S}^1$ using the interval $[-\pi,\pi)$. The discontinuity vanishes when using the angular distance $d_{\mathbb{S}^1}(\theta_1, \theta_2) = \lvert \operatorname{wrap}(\theta_1 - \theta_2) \rvert$. Therefore, we can define the Lipschitz constant with respect to $d_{\mathbb{S}^1}$.}. We use $m = 1\,\mathrm{kg}$, $l = 1\,\mathrm{m}$, $g = 10\,\mathrm{m/s^2}$, $b = 0$, $\tau = 0.1\,\mathrm{s}$, $\sigma_\theta = 0.01$, and $\sigma_\omega = 0.05$. The input set is $\mathcal{U} = \{-2, 0, 2\}\,\mathrm{N\cdot m}$. The input is not strong enough to lift the pendulum directly, so the controller has to perform swings.

\noindent\textbf{Abstraction.} The state space $\mathcal{X} = [-\pi, \pi] \times [-8, 8]$ is partitioned into $200 \times 200 = 40{,}000$ regions. The pendulum starts near the bottom at $x[0] = [0.99\pi, 0]^\top$ and must reach the goal set $\mathcal{G} = [-0.1\pi, 0.1\pi] \times [-1, 1]$.

\subsection{Mountain Car}
\textbf{System dynamics.} The Mountain Car models an underpowered vehicle in a valley. Its state is position and velocity: $x[k] = [p[k], v[k]]^\top$, and its input $u[k]$ is the throttle. The discrete-time dynamics are
\begin{subequations}
\begin{align*}
    \tilde{v}[k+1] &= \operatorname{clip}\big( v[k] + \tau (\alpha_{\mathrm{pwr}} u[k] - g_0 \cos(3 p[k])),\, -v_{\max},\, v_{\max} \big), \\
    p[k+1] &= p[k] + \tau\, \tilde{v}[k+1] + w_p[k], \\
    v[k+1] &= \tilde{v}[k+1] + w_v[k],
\end{align*}
\end{subequations}
where $\alpha_{\mathrm{pwr}}$ is the engine force coefficient, $g_0$ scales gravity, $w_p[k]$ and $w_v[k]$ are i.i.d.\ zero-mean Gaussian noise terms with standard deviations $\sigma_p$ and $\sigma_v$, and $\operatorname{clip}(z, \underline{z}, \bar{z}) \triangleq \max(\underline{z}, \min(z, \bar{z}))$. We use $\tau = 1\,\mathrm{s}$, $\alpha_{\mathrm{pwr}} = 0.0015$, $g_0 = 0.0025$, $v_{\max} = 0.07$, $\sigma_p = 10^{-4}$, and $\sigma_v = 10^{-5}$. The input set is $\mathcal{U} = \{-1, 0, 1\}$. The engine cannot climb the valley directly, so the car has to build momentum by moving back and forth. 

\noindent\textbf{Abstraction.}  The state space $\mathcal{X} = [-1.2, 0.6] \times [-0.07, 0.07]$ is partitioned into $500 \times 500 = 250{,}000$ regions. The car starts at the bottom of the valley, $x[0] = [-0.6, 0]^\top$, and must reach the goal set $\mathcal{G} = [0.45, 0.6] \times [0, 0.07]$.

\subsection{Unicycle}
\textbf{System dynamics.} The Unicycle moves in the 2D plane and must reach a goal region while avoiding obstacles. Its state is position and heading: $x[k] = [p_x[k], p_y[k], \theta[k]]^\top$. The inputs are the steering rate $u_1[k]$ and the velocity $u_2[k]$. $\alpha > 0$ is the steering coefficient. The discrete-time dynamics are
\begin{subequations}
\begin{align*}
    p_x[k+1] &= p_x[k] + \tau\, u_2[k] \cos\theta[k], \\
    p_y[k+1] &= p_y[k] + \tau\, u_2[k] \sin\theta[k], \\
    \theta[k+1] &= \operatorname{wrap}\left(\theta[k] + \tau \alpha\, u_1[k] + w_\theta[k]\right),
\end{align*}
\end{subequations}
where $w_\theta[k]$ is an i.i.d.\ zero-mean Gaussian noise term with standard deviation $\sigma_\theta = 0.05$. We use $\tau = 1\,\mathrm{s}$, $\alpha = 0.85$. The input set is $\mathcal{U} = \{-\tfrac{\pi}{2}, 0, \tfrac{\pi}{2}\} \times \{-1, 0, 1\}$.

\noindent\textbf{Abstraction.} The state space $\mathcal{X} = [0, 10]^2 \times [-\pi, \pi]$ is partitioned into $50 \times 50 \times 50 = 125{,}000$ regions. The vehicle starts in the lower-left corner at $x[0] = [0.5, 0.5, 0]^\top$ and must reach the goal set $\mathcal{G} = [8, 10] \times [0, 2] \times [-\pi, \pi]$ in the lower-right corner. It must avoid the unsafe set $[4.5, 5.5] \times [0, 7] \times [-\pi, \pi]$.

\subsection{Planar Robot}
\textbf{System dynamics.} The Planar Robot is moving on the 2D plane with friction. Its state is the position and velocity along each axis: $x[k] = [p_x[k], v_x[k], p_y[k], v_y[k]]^\top$. The inputs $u_1[k]$ and $u_2[k]$ are the accelerations along the $x$- and $y$-axes. $c_{\mathrm{d}} > 0$ is the friction coefficient. The discrete-time dynamics are
\begin{subequations}
\begin{align*}
    p_x[k+1] &= p_x[k] + \tau\, v_x[k] + \tfrac{\tau^2}{2} u_1[k] + w_{p_x}[k], \\
    v_x[k+1] &= \operatorname{clip}\big( v_x[k] + \tau\, u_1[k] - c_{\mathrm{d}}\, v_x[k] \lvert v_x[k] \rvert,\, v_{\min},\, v_{\max} \big), \\
    p_y[k+1] &= p_y[k] + \tau\, v_y[k] + \tfrac{\tau^2}{2} u_2[k] + w_{p_y}[k], \\
    v_y[k+1] &= \operatorname{clip}\big( v_y[k] + \tau\, u_2[k] - c_{\mathrm{d}}\, v_y[k] \lvert v_y[k] \rvert,\, v_{\min},\, v_{\max} \big),
\end{align*}
\end{subequations}
where $w_{p_x}[k]$ and $w_{p_y}[k]$ are i.i.d.\ zero-mean Gaussian noise terms with standard deviation $\sigma_p$, and $\operatorname{clip}(z, \underline{z}, \bar{z}) \triangleq \max(\underline{z}, \min(z, \bar{z}))$. We use $\tau = 1\,\mathrm{s}$, $c_{\mathrm{d}} = 0.1$, $v_{\min} = -2$, $v_{\max} = 2$, and $\sigma_p = 0.15$. The input set is $\mathcal{U} = \{-1, 0, 1\} \times \{-1, 0, 1\}$.

\noindent\textbf{Abstraction.} The state space $\mathcal{X} = \big([0, 10] \times [v_{\min}, v_{\max}]\big)^2$ is partitioned into $30^4 = 810{,}000$ regions. The robot starts near the origin at $x[0] = [0.01, 0.01, 0.01, 0.01]^\top$ and must reach the goal set $\mathcal{G} = \big([7, 10] \times [v_{\min}, v_{\max}]\big)^2$.

\begin{table}[t]
\centering
\caption{Memory usage (MB) of the \gls{fimdp} and \gls{imdp} abstractions. Both implementations include sparsity.}
\label{tab:memory}
\setlength{\tabcolsep}{4pt}
\begin{tabular}{l ccc}
\toprule
 & \gls{imdp} & \gls{fimdp} & \\
System & \cite{nazeri2025data2} & (ours) & Reduction \\
\midrule
Double integrator & 32.09       & \textbf{11.04}      & $2.9\times$  \\
Pendulum          & 162.05      & \textbf{35.25}      & $4.6\times$  \\
Mountain car      & 369.07      & \textbf{140.19}     & $2.6\times$  \\
Unicycle          & 6{,}196.98  & \textbf{401.26}     & $15.4\times$ \\
Planar Robot             & 80{,}201.57 & \textbf{3{,}392.72} & $23.6\times$ \\
\bottomrule
\end{tabular}
\end{table}

\begin{table}[t]
\centering
\caption{Satisfaction probability and computation time for different benchmarks and sampling sizes. Our \gls{fimdp} approach is consistently less conservative and more efficient. }
\label{tab:prob_time}
\setlength{\tabcolsep}{4pt}
{\small\begin{tabular}{l c r cc cc}
\toprule
 & & & \multicolumn{2}{c}{Satisfaction prob.} & \multicolumn{2}{c}{Time (s)} \\
\cmidrule(lr){4-5} \cmidrule(lr){6-7}
 & & & \gls{imdp} & \gls{fimdp} & \gls{imdp} & \gls{fimdp} \\
System & $n$ & $N$ & \cite{nazeri2025data2} & (ours) & \cite{nazeri2025data2} & (ours) \\
\midrule
\multirow{3}{*}{Double integrator} & \multirow{3}{*}{2}
 & $1{,}000$  & 90.14 & \textbf{91.76} & 4.1  & \textbf{3.2}  \\
 & & $5{,}000$  & 98.16 & \textbf{98.47} & 10.0 & \textbf{8.2}  \\
 & & $25{,}000$ & 99.63 & \textbf{99.70} & 37.1 & \textbf{35.8} \\
\midrule
\multirow{3}{*}{Pendulum} & \multirow{3}{*}{2}
 & $1{,}000$  & 34.97 & \textbf{49.41} & 9.9   & \textbf{7.0}  \\
 & & $5{,}000$  & 77.73 & \textbf{81.26} & 23.4  & \textbf{19.9} \\
 & & $25{,}000$ & 89.27 & \textbf{90.84} & 103.6 & \textbf{96.9} \\
\midrule
\multirow{3}{*}{Mountain car} & \multirow{3}{*}{2}
 & $1{,}000$  & 7.38  & \textbf{9.35}  & 47.0  & \textbf{41.2}  \\
 & & $5{,}000$  & 59.37 & \textbf{62.62} & 131.7 & \textbf{126.0} \\
 & & $25{,}000$ & 88.79 & \textbf{90.17} & 610.2 & \textbf{558.4} \\
\midrule
\multirow{3}{*}{Unicycle} & \multirow{3}{*}{3}
 & $1{,}000$  & 0.09  & \textbf{26.80} & 154.8     & \textbf{58.6}      \\
 & & $5{,}000$  & 56.27 & \textbf{75.99} & 371.2     & \textbf{192.4}     \\
 & & $25{,}000$ & 86.06 & \textbf{92.19} & 1{,}810.1 & \textbf{1{,}017.1} \\
\midrule
\multirow{3}{*}{Planar Robot} & \multirow{3}{*}{4}
 & $1{,}000$  & OOM & \textbf{50.15} & OOM & \textbf{332.9}     \\
 & & $5{,}000$  & OOM & \textbf{78.80} & OOM & \textbf{1{,}520.8} \\
 & & $25{,}000$ & OOM & \textbf{86.84} & OOM & \textbf{8{,}220.9} \\
\bottomrule
\end{tabular}}
\end{table}

\section{Conclusion \& Future Work}
\label{sec:conclusion}
We introduced a data-driven abstraction framework for \glspl{fimdp} that leverages factorization to combat the pervasive challenge of state-space explosion. Numerical experiments show that compared to the state-of-the-art, the framework provides tighter satisfaction probability guarantees, enables faster construction of the abstraction, and reduces memory consumption.
Moreover, it scales to a 4D system, which was previously intractable for abstraction-based data-driven verification. 

Under the assumption of an orthogonal transition kernel on the concrete system, the framework provides statistical guarantees that the behaviours of the concrete system are captured by the constructed abstract model. Future directions may explore approaches to relax this assumption while preserving the benefits of the factorization. 

A theoretical analysis of the sample complexity reveals how the framework is able to obtain tighter bounds on the transition probabilities and, by extension, the satisfaction probabilities. Moreover, the analysis hints at directions that may further reduce the sample complexity, including online adaptive partitioning and dependency graph detection.

\begin{acks}
This research has been supported by EPSRC grant EP/Y028872/1, Mathematical Foundations of Intelligence: An “Erlangen Programme” for AI.
\end{acks}

\bibliographystyle{ACM-Reference-Format}
\bibliography{references}

\appendix

\section{The necessity of a decomposed transition kernel}\label{app:decomposition}
Assumption~\ref{ass:independent_noise} restricts the class of systems for which the method is applicable.
Therefore, it is desirable to know if the condition is necessary or only sufficient to enable the abstraction of the system to an \gls{fimdp}.
First, we restrict ourselves to \glspl{fimdp} with $n$ components; that is, a non-trivial factorization. 
Next, we in fact relax \Cref{ass:independent_noise}. Instead, we prove the result if the corresponding transition kernel can be decomposed on given grid.
To this end, assume that $\regionset_i$ and $\stateset_i$ are given for each $i \in \{1, \ldots, n\}$ as defined in \Cref{subsec:abstraction_process}. 
We define the inverse projection $(\projection_i)^{-1} : \regionset_i \to 2^{\reals^n}$ for a Borel measurable set $\region_i \in \regionset_i$ as the slice of $\reals^n$ corresponding to $\region_i$. That is,
\begin{equation}
    (\projection_i)^{-1}(\region_i) = \reals^{i - 1} \times \region_i \times \reals^{n - i}.
\end{equation}
Assume that a set of actions $\actionset$ is defined with an associated control input $\overline{\control}_\action$ for each $\action \in \actionset$.
A transition kernel is factorizable into $n$ components \emph{on the grid defined by $\regionset$} if and only if there exists marginal transition kernels $\transitionkernel_i$, $i \in \{1, \ldots, n\}$ such that the following holds for all $\source \in \stateset$, $\action \in \actionset$, and $\sysstate \in \region_\source$,
\begin{equation}\label{eq:grid_factorized_transition_kernel}
    \transitionkernel(\region_{\destination} \mid \sysstate, \overline{\control}_\action) = \prod_{i=1}^n \transitionkernel_i(\region_{\destination_i} \mid \sysstate, \overline{\control}_\action) \qquad \text{for all } \destination \in \stateset.
\end{equation}
The goal is to show that \eqref{eq:matching} can only hold if \eqref{eq:grid_factorized_transition_kernel} holds.

By the structure of the \gls{fimdp}, each ambiguity set $\ambiguityset_{\source,\action}$ is decomposed into marginals:
\begin{equation}\label{eq:product-ambiguity}
    \ambiguityset_{\source,\action} = \left\{ \probdist^1_{\source,\action} \otimes \cdots \otimes \probdist^n_{\source,\action} \;:\; \probdist^i_{\source,\action} \in \ambiguityset^l_{\source,\action},\ i = 1, \ldots, n \right\},
\end{equation}
where $\ambiguityset^l_{\source,\action}$ is a set of distributions over $\stateset_i$ and $(\probdist^1_{\source,\action} \otimes \cdots \otimes \probdist^n_{\source,\action})(\destination) = \prod_{i=1}^n \probdist^i_{\source,\action}(\destination^l)$. Consequently, for \eqref{eq:matching} to hold, it requires that
\begin{equation}\label{eq:transition_kernel_equal_factored_ambiguity_set}
    \transitionkernel(\region_{\destination} \mid \sysstate, \overline{\control}_\action) = \prod_{i=1}^n \probdist^i_{\source,\action}(\destination_i) \qquad \text{for all } \destination \in \stateset.
\end{equation}
We show that the equality in \eqref{eq:transition_kernel_equal_factored_ambiguity_set} can only hold if \eqref{eq:grid_factorized_transition_kernel} holds. 

\begin{proposition}[Necessity of decomposition]\label{prop:necessity}
    Fix $\source \in \stateset$ and an action $\action$, and let $\ambiguityset_{\source,\action}$ be of the form~\eqref{eq:product-ambiguity}. For every $\sysstate \in \region_\source$, if~\eqref{eq:matching} holds with some $\probdist_{\source,\action} = \probdist^1_{\source,\action} \otimes \cdots \otimes \probdist^n_{\source,\action} \in \ambiguityset_{\source,\action}$, then, there exists marginal transition kernels $\transitionkernel_i$, $i = 1, \ldots, n$ such that \eqref{eq:grid_factorized_transition_kernel} holds for all $\sysstate \in \region_\source$ and for all $i \in \{1, \ldots, n\}$ and $\destination_i \in \stateset_i$,
    \begin{equation}\label{eq:marginal-identification}
        \probdist^i_{\source,\action}(\destination_i) = \transitionkernel_i(\region_{\destination_i} \mid \sysstate, \overline{\control}_\action).
    \end{equation}
\end{proposition}

\begin{proof}
    Fix $\sysstate \in \region_{\source}$ and suppose that~\eqref{eq:matching} holds with some product distribution
    $\probdist_{\source,\action} = \probdist^1_{\source,\action} \otimes \cdots \otimes \probdist^n_{\source,\action} \in \ambiguityset_{\source,\action}$.
    Fix $i \in \{1,\ldots,n\}$ and $\destination_i \in \stateset_i$. Write $\stateset_{-i} = \prod_{j \neq i} \stateset_j$, and for $\destination_{-j} \in \stateset_{-j}$ let $(\destination_i, \destination_{-i}) \in \stateset$ denote the joint abstract state with $i$-th component $\destination_i$.

    Consider the family of Borel measurable (potentially unbounded) hyperrectangles $\{\region_{(\destination_i, \destination_{-i})}\}_{\destination_{-i} \in \stateset_{-i}}$.

    \emph{(i) Pairwise disjointness.} If $\destination_{-i} \neq \tilde{\destination}_{-i}$, then there exists $j \neq i$ with $\destination_j \neq \tilde{\destination}_j$. Hence, $\region_{\destination_j} \cap \region_{\tilde{\destination}_j} = \emptyset$, which implies $\region_{(\destination_i, \destination_{-i})} \cap \region_{(\destination_i, \tilde{\destination}_{-i})} = \emptyset$.

    \emph{(ii) Covering.} Using distributivity of products over unions and that $\{\region_{\destination_j}\}_{\destination_j \in \stateset_j}$ partitions $\reals$ for each $j$, we have
    \begin{equation}
        \bigcup_{\destination_{-i} \in \stateset_{-i}} \region_{(\destination_i, \destination_{-i})} 
            = \region_{\destination_i} \times \prod_{j \neq i} \left(\bigcup_{\destination_j \in \stateset_j} \region_{\destination_j}\right)
            = (\projection_i)^{-1}(\region_{\destination_i}).
    \end{equation}

    We define the $i$-th marginal of $\transitionkernel$ as the pushforward
    \begin{equation}
        \transitionkernel_i(B_i \mid \sysstate, \control) \coloneqq \transitionkernel\left((\projection_i)^{-1}(B_i) \mid \sysstate, \control\right)
    \end{equation}
    for any Borel set $B_i \subseteq \reals$. By the definition of the pushforward marginal kernel, (i)-(ii), countable additivity of $\transitionkernel(\cdot\mid \sysstate,\overline{\control}_\action)$, and~\eqref{eq:matching},
    \begin{equation}
        \begin{aligned}
            \transitionkernel_i(\region_{\destination_i} \mid \sysstate, \overline{\control}_\action)
            &= \transitionkernel\left((\projection_i)^{-1}(\region_{\destination_i}) \mid \sysstate, \overline{\control}_\action\right)\\
            &= \sum_{\destination_{-i} \in \stateset_{-i}} \transitionkernel\left(\region_{(\destination_i,\destination_{-i})} \mid \sysstate, \overline{\control}_\action\right)\\
            &= \sum_{\destination_{-i} \in \stateset_{-i}} \probdist^i_{\source,\action}(\destination_i) \prod_{j \neq i} \probdist^j_{\source,\action}(\destination_j)\\
            &= \probdist^i_{\source,\action}(\destination_i) \prod_{j \neq i} \left(\sum_{\destination_j \in \stateset_j} \probdist^j_{\source,\action}(\destination_j)\right)
            = \probdist^i_{\source,\action}(\destination_i),
        \end{aligned}
    \end{equation}
    since each $\probdist^j_{\source,\action}$ is a probability distribution on $\stateset_j$. This proves~\eqref{eq:marginal-identification}.
    Substituting~\eqref{eq:marginal-identification} into~\eqref{eq:matching} yields, for all $\destination \in \stateset$,
    \begin{equation}
        \transitionkernel(\region_{\destination} \mid \sysstate, \overline{\control}_\action)
        = \prod_{i=1}^n \probdist^i_{\source,\action}(\destination^l)
        = \prod_{i=1}^n \transitionkernel_i(\region_{\destination_i} \mid \sysstate, \overline{\control}_\action),
    \end{equation}
    which is~\eqref{eq:grid_factorized_transition_kernel}.
\end{proof}

Proposition~\ref{prop:necessity} shows that a product-form ambiguity set cannot soundly abstract a system whose transition kernel does not factorize over the axes, regardless of how large the marginal sets $\ambiguityset^l_{\source,\action}$ are chosen to be.

\begin{example}[Factoriable kernel only on a grid]
    It is possible for the transition kernel to be unfactorizable, i.e., the axes are dependent, outside of a given gridding. As an example, consider two independent fair Bernoulli variables $B_1,B_2$ and let $Z$ be an independent uniform variable on $(0, 1)$. Define $Y_1 = 2B_1 + Z$ and $Y_2 = 2B_2 + Z$, and partition each axis into $[0, 1)$, $[2, 3)$ and the complement. The cell indices are independent, so their joint probabilities have product form, but the continuous coordinates are dependent through their identical fractional parts.
\end{example}

\section{Trivial and Non-Trivial Specifications}
\label{app:trivial}

Data-driven abstractions are built from a finite set of samples.
A finite number of samples can never rule out a transition, so the concentration bounds assign a positive upper bound $\overline{\probdist}_{\source,\action}(\destination) > 0$ for each triplet $(\source,\action,\destination) \in \stateset \times \actionset \times \stateset$, including those never observed, while the corresponding lower bound $\underline{\probdist}_{\source,\action}(\destination)$ is $0$ if the transition is never observed. Thus, for the purpose of this analysis, we make the following assumption
\begin{assumption}[Upper-bound Positivity]\label{ass:positivity}
    For all $\source \in \stateset$, $\action \in \actionset$, and $\destination \in \stateset$, there exists $\probdist_{\source,\action} \in \ambiguityset_{\source,\action}$ with $\probdist_{\source,\action}(\destination) > 0$.
\end{assumption}
An environment strategy can therefore always inject a small amount of probability mass into \emph{every} abstract state, at every step. In other words, the support of $\ambiguityset_{\source,\action}$ is full.
Over an infinite horizon this repeated leakage dominates the analysis.
For example, the robust (pessimistic) probability of an invariance specification $\mathsf{G}\,\mathit{safe}$ is $0$ in every state, for every controller and every amount of data, even if the concrete system satisfies it with probability close to $1$.
Such bounds are artefacts of the abstraction rather than properties of the system: they carry no information, and controller synthesis against them returns an arbitrary controller.
It is therefore important to know \emph{a priori} which specifications can yield informative bounds.
This appendix gives an exact classification.

We call a specification \emph{small} if its pessimistic value is forced to $0$, by the positivity of the upper bounds alone. We can similarly say that a specification is \emph{large} if its optimistic value is forced to $1$; however, since we are only interested in a lower bound, we restrict the analysis to small specifications. The analysis for large specifications follows analogously. 
Due to the full support of the system model, the classification depends only on the specification and the set of labels used by the abstraction, and it can be computed in time linear in the size of the automaton.
One approach to avoid full support is to provide more information about the system. In particular, it is enough to have (conservative) bounds on the support of the noise, while the noise distribution and dynamics are still inferred form data.

\subsection{Reachable Bottom SCCs}

Let $\specification$ be an $\omega$-regular specification over $\outputspace$ in positive normal form, and let $\hat{\specification}$ denote the specification over $\hat{\outputspace}$ obtained by replacing each $\neg\outputvar$ with $\overline{\outputvar}$.
Let $\automaton = (\automatonstatespace, \alphabet, \automatontransition, \automatonstate_0, \dpapriority)$ be a \gls{dpa} for $\hat{\specification}$ over the alphabet $\alphabet = 2^{\hat{\outputspace}}$, in which $\outputvar$ and $\overline{\outputvar}$ are treated as independent propositions, and let $n = |\automatonstatespace|$. 
Only the letters that are actually emitted by the abstraction matter.
 
\begin{definition}[Label graph]
Let $\alphabet_\labelling = \{ \labelling(\source) : \source \in \stateset \} \subseteq 2^{\hat{\outputspace}}$ be the set of emitted letters.
The \emph{label graph} of $\automaton$ is the directed graph $G_\labelling = (\automatonstatespace, E_\labelling)$ with
$E_\labelling = \{ (\automatonstate, \automatontransition(\automatonstate, \omegaletter)) : \automatonstate \in \automatonstatespace,\ \omegaletter \in \alphabet_\labelling \}$.
For $\initialsource \in \stateset$, the \emph{initial automaton state} is $\automatonstate^{\initialsource} = \automatontransition(\automatonstate_0, \labelling(\initialsource))$, and $\mathrm{Reach}_\labelling(\initialsource) \subseteq \automatonstatespace$ is the set of states reachable from $\automatonstate^{\initialsource}$ in $G_\labelling$.
\end{definition}

\begin{definition}[Bottom SCCs]
A \gls{scc} of $G_\labelling$ is a maximal set $C \subseteq \automatonstatespace$ such that every state of $C$ is reachable from every other state of $C$.
An \gls{scc} $C$ is a \gls{bscc} if it has no outgoing edges, i.e., $\automatontransition(\automatonstate, \omegaletter) \in C$ for all $\automatonstate \in C$ and $\omegaletter \in \alphabet_\labelling$.
The set of \emph{reachable \glspl{bscc}} from $\initialsource$ is
$\mathcal{B}_\labelling(\initialsource) = \{ C \text{ \acrshort{bscc} of } G_\labelling : C \subseteq \mathrm{Reach}_\labelling(\initialsource) \}$.
A \gls{bscc} $C$ is \emph{accepting} if $\min_{\automatonstate \in C} \mathsf{pr}(\automatonstate)$ is even, and \emph{rejecting} otherwise.
\end{definition}
 
Since $\automatontransition$ is total and $\alphabet_\labelling \neq \emptyset$, every state in $G_\labelling$ has a successor.
Hence, every \gls{bscc} contains a cycle, every state may reach some \gls{bscc} along a path of length at most $n$, and $\mathcal{B}_\labelling(\initialsource) \neq \emptyset$.
 
\begin{definition}[Small]
\label{def:small-large}
The specification $\hat{\specification}$ is \emph{small} from $\initialsource$ if every $C \in \mathcal{B}_\labelling(\initialsource)$ is rejecting.
\end{definition}

The notion of a small specification depends on the abstraction only through $\alphabet_\labelling$ and $\labelling(\initialsource)$, and not on its transitions, actions, or interval values, as the distribution is full.
Under Assumption~\ref{ass:positivity}, the computation and labelling of \glspl{bscc} can performed on $\automaton$ alone:
\begin{enumerate}
  \item Compute $\alphabet_\labelling$ in $O(|\stateset| \cdot |\outputspace|)$ time.
  \item Construct $G_\labelling$, with $|\automatonstatespace| \cdot |\alphabet_\labelling|$ edges.
  \item Compute the \glspl{scc} of $G_\labelling$ with Tarjan's algorithm, and mark those without outgoing edges  as bottom.
  \item Label each \gls{bscc} accepting or rejecting via its minimal priority.
  \item By backward search in $G_\labelling$, compute the set $R_{\mathrm{acc}}$ of states that can reach an accepting \gls{bscc}.
\end{enumerate}
Then $\hat{\specification}$ is small from $\initialsource$ if and only if $\automatonstate^{\initialsource} \notin R_{\mathrm{acc}}$.
Determinism of $\automaton$ is essential: it ensures that each label sequence has a unique run, so that the \gls{bscc} in which the run settles determines acceptance.

\subsection{Classification Results}

Fix an initial state $\source_0 \in \stateset$.
For a scheduler $\scheduler \in \schedulerset_M$ and an environment strategy $\adversary$, let $\probmeas^{M,\scheduler,\adversary}_{\source_0}$ be the induced measure on paths $\source_0 \action_0 \source_1 \action_1 \ldots$, as in \Cref{subsec:fimdp}.
Let $\mathcal{F}_k$ denote the $\sigma$-algebra generated by $\source_0, \action_0, \ldots, \source_k, \action_k$.
Since the environment selects the distribution of $\source_{k+1}$ from $\ambiguityset_{\source_k,\action_k}$ after observing the history up to $\action_k$, the conditional distribution $\probmeas^{M,\scheduler,\adversary}_{\source_0}(\source_{k+1} = \cdot \mid \mathcal{F}_k)$ is the distribution selected at step $k$.
 
\begin{definition}[$\varepsilon$-mixing]
For $\varepsilon > 0$, an environment strategy $\adversary$ is \emph{$\varepsilon$-mixing} if every distribution it selects assigns probability at least $\varepsilon$ to every state in $\stateset$.
\end{definition}
 
\begin{lemma}[Interior distribution]
\label{lem:interior}
Under Assumption~\ref{ass:positivity}, there exist $\varepsilon > 0$ and, for every $(\source,\action)$, a distribution $\gamma^*_{\source,\action} \in \ambiguityset_{\source,\action}$ with $\probdist^\star_{\source,\action}(\destination) \ge \varepsilon$ for all $\destination \in \stateset$.
Consequently, the environment strategy $\adversary^\star$ that selects $\probdist^\star_{\source_k,\action_k}$ at every step $k$ is $\varepsilon$-mixing.
\end{lemma}
 
\begin{proof}
Fix $(\source,\action)$ and $i$.
For each $\destination_i \in \stateset_i$, Assumption~\ref{ass:positivity} provides $\probdist_{\source,\action}^{i,\destination} \in \ambiguityset^i_{\source,\action}$ with $\probdist^{i,\destination}(\destination) > 0$.
Since $\ambiguityset^i_{\source,\action} \in \intamb(\stateset_i)$ is convex, the average $\probdist^{i,\star} = \frac{1}{|\stateset_i|} \sum_{\destination_i \in \stateset_i} \probdist_{\source,\action}^{i,\destination}$ lies in $\ambiguityset^i_{\source,\action}$ and is positive on all of $\stateset_i$.
By Definition~\ref{def:fimdp}, the product $\probdist^\star_{\source,\action} = \bigotimes_{i=1}^n \probdist_{\source,\action}^{i,\star}$ lies in $\ambiguityset_{\source,\action}$.
It satisfies $\probdist^\star_{\source,\action}(\destination) = \prod_{i=1}^n \probdist^{i,\star}(\destination_i) > 0$ for every $\destination = (\destination_1, \ldots, \destination_n) \in \stateset$.
Let $\varepsilon$ be the minimum of these finitely many positive values.
\end{proof}
 
Let $\rho = \automatonstate_0 \automatonstate_1 \ldots$ denote the run of $\automaton$ on $\labelling(\source_0) \labelling(\source_1) \ldots$, so that $\automatonstate_1 = \automatonstate^{\source_0}$ and $\automatonstate_{k+1} = \automatontransition(\automatonstate_k, \labelling(\source_k))$.
Note that $\automatonstate_{k+1}$ is $\mathcal{F}_k$-measurable.
 
\begin{lemma}[Runs under mixing environments]
\label{lem:mixing}
Let $\scheduler$ be any scheduler and $\adversary$ an $\varepsilon$-mixing environment strategy.
Then $\probmeas^{M,\scheduler,\adversary}_{\source_0}$-almost surely:
(a) the run $\rho$ eventually enters some $C \in \mathcal{B}_\labelling(\source_0)$ and remains in it; and
(b) $\mathrm{Inf}(\rho) = C$.
\end{lemma}
 
\begin{proof}
Write $\probmeas = \probmeas^{M,\scheduler,\adversary}_{\source_0}$.
Since $\adversary$ is $\varepsilon$-mixing, $\probmeas(\source_{k+1} = \destination \mid \mathcal{F}_k) \ge \varepsilon$ almost surely for all $k$ and $\destination$.
By the tower property, for all $k \ge 0$ and $\destination_1, \ldots, \destination_\ell \in \stateset$,
$\probmeas(\source_{k+1} = \destination_1, \ldots, \source_{k+\ell} = \destination_\ell \mid \mathcal{F}_k) \ge \varepsilon^\ell$ almost surely.
For each $\omegaletter \in \alphabet_\labelling$, fix $\source_\omegaletter \in \stateset$ with $\labelling(\source_\omegaletter) = \omegaletter$.
Every edge $(\automatonstate, \automatontransition(\automatonstate,\omegaletter))$ of $G_\labelling$ is taken at step $k+1$ if $\source_{k+1} = \source_\sigma$.
Hence, for any path of length $\ell$ in $G_\labelling$ starting at $\automatonstate_{k+1}$, the run follows it during steps $k+2, \ldots, k+\ell+1$ with conditional probability at least $\varepsilon^\ell$ given $\mathcal{F}_k$.
 
(a) Let $\automatonstatespace_B$ be the union of all \glspl{bscc} of $G_\labelling$.
Every state \emph{can} reach $\automatonstatespace_B$ along some path of length at most $|\automatonstatespace|$, and $Z_B$ is closed under the edges of $G_\labelling$.
For $r \in \naturals$, let $t_r = 1 + r|\automatonstatespace|$ denote the start of the $r$-th block of $|\automatonstatespace|$ steps.
On the event $\{\automatonstate_{t_r} \notin Z_B\}$, the run enters $Z_B$ by time $t_{r+1}$ with conditional probability at least $\varepsilon^{|\automatonstatespace|}$ given $\mathcal{F}_{t_r - 1}$.
By induction, $\probmeas(\automatonstate_{t_r} \notin \automatonstatespace_B) \le (1 - \varepsilon^{|\automatonstatespace|})^r \to 0$.
Once entered, $Z_B$ is never left, and the run stays in the \gls{bscc} $C$ it enters.
Since all letters $\labelling(\source_k)$ lie in $\alphabet_\labelling$, the run moves along edges of $G_\labelling$ from $\automatonstate_1 = \automatonstate^{\source_0}$, so $C \in \mathcal{B}_\labelling(\source_0)$.
 
(b) Fix $C \in \mathcal{B}_\labelling(\source_0)$, $\automatonstate' \in C$ and $k_0 \ge 1$.
Since $C$ is strongly connected and contains a cycle, every state of $C$ reaches $\automatonstate'$ by a path in $C$ of length between $1$ and $|\automatonstatespace|$.
Repeating the block argument of (a) from time $k_0$ shows that the event $\{\automatonstate_{k_0} \in C\} \cap \{\automatonstate_k \neq \automatonstate' \text{ for all } k > k_0\}$ has probability $0$.
Taking the countable union over $k_0$ and the finite union over $(C, \automatonstate')$, almost surely every state of the entered \gls{bscc} is visited infinitely often, so $\mathrm{Inf}(\rho) \supseteq C$.
Together with (a), $\mathrm{Inf}(\rho) = C$.
\end{proof}
 
Since acceptance of $\rho$ is determined by $\min_{\automatonstate \in \mathrm{Inf}(\rho)} \dpapriority(\automatonstate)$, Lemma~\ref{lem:mixing} implies that, under any $\varepsilon$-mixing environment, $\probmeas^{M,\scheduler,\adversary}_{\initialsource}(\specification)$ equals the probability that the run settles in an accepting \gls{bscc}.
 
\begin{theorem}[Forced triviality]
\label{thm:trivial}
Under Assumption~\ref{ass:positivity}, if $\specification$ is small from $\initialsource$, then $\probmeas^{M,\scheduler}_{\initialsource}(\specification) = 0$ for every scheduler $\scheduler$.
In particular, $\sup_{\scheduler \in \schedulerset_M} \probmeas^{M,\scheduler}_{\initialsource}(\specification) = 0$.
\end{theorem}
 
\begin{proof}
Let $\scheduler$ be any scheduler, and let $\adversary^*$ be the $\varepsilon$-mixing environment strategy of Lemma~\ref{lem:interior}.
By Lemma~\ref{lem:mixing}, the run almost surely settles in some $C \in \mathcal{B}_\labelling(\initialsource)$ with $\mathrm{Inf}(\rho) = C$, and every such $C$ is rejecting.
Hence $\probmeas^{M,\scheduler}_{\initialsource}(\specification) \le \probmeas^{M,\scheduler,\adversary^*}_{\initialsource}(\specification) = 0$.
\end{proof}
 
The classification is exact: smallness characterises precisely the specifications for which a robust satisfaction probability of $0$ is forced by Assumption~\ref{ass:positivity}.
 
\begin{theorem}[Tightness]
\label{thm:tight}
Fix $\stateset_1, \ldots, \stateset_n$, $\actionset$ and $\labelling$.
If $\specification$ is not small from $\initialsource$, there exists an \gls{fimdp} $M$ over these sets satisfying Assumption~\ref{ass:positivity} with $\probmeas^{M,\scheduler}_{\initialsource}(\specification) > 0$ for every scheduler $\scheduler$.
\end{theorem}
 
\begin{proof}
For all $\source$, $\action$ and $i$, let $\ambiguityset^i_{\source,\action} = \{\mathcal{U}(\stateset_i)\}$, where $\mathcal{U}(\stateset_i)$ is the uniform distribution on $\stateset_i$.
This is an interval ambiguity set with both lower and upper bounds equal to $\frac{1}{|\stateset_i|}$, and it satisfies Assumption~\ref{ass:positivity}.
Then $\ambiguityset_{\source,\action}$ contains only the uniform distribution on $\stateset$.
Hence, every environment strategy selects it at every step and is $|\frac{1}{|\stateset|}$-mixing.
Since $\specification$ is not small, some accepting $C \in \mathcal{B}_\labelling(\initialsource)$ is reachable from $\automatonstate^{\initialsource}$ by a path of length $\ell \le |\automatonstatespace|$.
The run follows this path with probability at least $\frac{1}{|\stateset|^{\ell}}$, then remains in $C$, and by Lemma~\ref{lem:mixing}(b) satisfies $\mathrm{Inf}(\rho) = C$ almost surely.
Hence $\probmeas^{M,\scheduler,\adversary}_{\initialsource}(\specification) \ge \frac{1}{|\stateset|^{|\automatonstatespace|}}$ for all $\scheduler$ and $\adversary$, so $\probmeas^{M,\scheduler}_{\initialsource}(\specification) \ge \frac{1}{|\stateset|^{|\automatonstatespace|}} > 0$.
\end{proof}

\begin{remark}[Scope]
Non-smallness does not imply $\probmeas^{M,\scheduler}_{\source_0}(\specification) > 0$ for a \emph{given} \gls{fimdp} and scheduler.
For instance, zero lower bounds may allow the environment to avoid a target set forever.
Theorem~\ref{thm:tight} only shows that the triviality is not forced by Assumption~\ref{ass:positivity}.
Moreover, the collapse in Theorem~\ref{thm:trivial} concerns the infinite horizon only: bounded-horizon variants of small specifications generally have non-trivial values.
\end{remark}

\end{document}